%% file: adsbranefinal5.tex
\documentclass[12pt]{article}
\usepackage{latexsym}
\usepackage{epsfig,amssymb,euscript}
\usepackage{amsmath}
\usepackage{array,calc,epsfig}
\usepackage[linktocpage]{hyperref}
\usepackage{pstricks}
\usepackage{bbm}
\usepackage{fancybox}

\def\be{\begin{equation}}
\def\ee{\end{equation}}
\newcommand{\eq}[1]{\begin{equation}#1\end{equation}}
\def\bea{\begin{eqnarray}}
\def\eea{\end{eqnarray}}
\newcommand\bbone{\ensuremath{\mathbbm{1}}}

\numberwithin{equation}{section} 
\usepackage[]{graphicx}
\def\cala         {{\cal A}}

\def\calb         {{\cal B}}

\def\cald         {{\cal D}}
\def\cale         {{\cal E}}
\def\calf         {{\cal F}}
\def\calg         {{\cal G}}

\def\calk         {{\cal K}}
\def\call         {{\cal L}}
\def\calm         {{\cal M}}
\def\caln         {{\cal N}}

\def\calv         {{\cal V}}
\def\calw         {{\cal W}}

\def\Re           {{\rm Re\hskip0.1em}}
\def\Im           {{\rm Im\hskip0.1em}}

\def\sqr#1#2{{\vcenter{\vbox{\hrule height.#2pt
 \hbox{\vrule width.#2pt height#1pt \kern#1pt \vrule width.#2pt}\hrule
 height.#2pt}}}}

\def\d{\text{d}}

\def\slashchar#1{\setbox0=\hbox{$#1$}           
\dimen0=\wd0                                 
\setbox1=\hbox{/} \dimen1=\wd1               
\ifdim\dimen0>\dimen1                        
\rlap{\hbox to \dimen0{\hfil/\hfil}}      
#1                                        
\else                                        
\rlap{\hbox to \dimen1{\hfil$#1$\hfil}}   
/                                         
\fi}

\begin{document}
\font\cmss=cmss10 \font\cmsss=cmss10 at 7pt

\vskip -0.5cm
\rightline{\small{\tt MPP-2007-162}}
\rightline{\small{\tt LMU-ASC 69/07}}
\rightline{\small{\tt arXiv:0710.5530}}

\vskip .7 cm

\hfill
\vspace{18pt}
\begin{center}
{\Large \textbf{D-branes on AdS flux compactifications}}
\end{center}

\vspace{6pt}
\begin{center}
{\large\textsl{Paul Koerber~$^{a}$ and Luca Martucci~$^{b}$}}

\vspace{25pt}
\textit{\small $^a$ Max-Planck-Institut f\"{u}r Physik -- Theorie,\\
                    F\"{o}hringer Ring 6,  D-80805 M\"{u}nchen, Germany}\\ \vspace{6pt}
\textit{\small $^b$ Arnold Sommerfeld Center for Theoretical Physics,\\ LMU M\"unchen,
Theresienstra\ss e 37, D-80333 M\"unchen, Germany}\\  \vspace{6pt}
\end{center}

\vspace{12pt}

\begin{center}
\textbf{Abstract}
\end{center}

\vspace{4pt} {\small
We study D-branes in $\caln=1$ flux compactifications to AdS$_4$. We derive their supersymmetry conditions and express them in terms of background generalized calibrations. Basically because AdS has a boundary, the analysis of stability is more subtle and
qualitatively different from the usual case of Minkowski compactifications. For instance, stable D-branes filling AdS$_4$ may wrap trivial internal cycles. Our analysis gives a geometric realization of the four-dimensional field theory approach of Freedman and collaborators. Furthermore, the one-to-one correspondence between the supersymmetry conditions of the background and the existence of generalized calibrations for D-branes is clarified and extended to any supersymmetric flux background that admits a time-like Killing vector and for which all fields are time-independent with respect to the associated time. As explicit examples, we discuss supersymmetric D-branes on IIA nearly K\"ahler AdS$_4$  flux compactifications.

\noindent }

\vspace{1cm}


\thispagestyle{empty}

\vfill
\vskip 5.mm
\hrule width 5.cm
\vskip 2.mm
{\small
\noindent e-mail: koerber@mppmu.mpg.de, luca.martucci@physik.uni-muenchen.de
}

\newpage

\setcounter{footnote}{0}

\tableofcontents

\newpage


\section{Introduction and summary}

Supersymmetric compactifications to AdS$_4$ have some qualitatively different properties
from compactifications to Minkowski space. The supersymmetry conditions seem more restrictive
in the AdS case. While in the Minkowski case fluxes are added for reasons such
as moduli stabilization, they are unavoidable in the AdS case. Furthermore, classically\footnote{Starting from a classical $SU(3)$-structure compactification in IIB, AdS vacua are possible after taking into account non-perturbative corrections, as in the example of \cite{kklt}. The internal geometry of these kinds of vacua has been discussed in \cite{effective},
where it has been shown how non-perturbative corrections arising from localized D-instantons can destabilize the $SU(3)$-structure into a general $SU(3)\times SU(3)$-structure,
while the $SU(3)$-structure can be preserved only if the D-instantons are smeared in the internal space.} an
$SU(3)$-structure compactification is only possible in type IIA supergravity, while for type IIB
we must have a static $SU(2)$-structure or, presumable, a more
general $SU(3)\times SU(3)$-compactification, although examples of the latter are not known yet.
The type IIA $SU(3)$-structure compactifications are phenomenologically interesting as it is possible
in this setting to construct models with all moduli fixed (see e.g. \cite{zwirner,dewolfe,palti,camara,acharya}).

As shown in \cite{gmpt}, also for AdS compactifications the supersymmetry conditions are naturally expressed in the language of
generalized complex (GC) geometry \cite{hitchin,gualtieri} in terms of two compatible pure spinors of $SO(6,6)$: $\Psi_1$ and $\Psi_2$.
They can also be thought of as polyforms, sums of forms of different even/odd dimensions, and
it may be useful to keep in mind the $SU(3)$-structure case where they correspond to $\Omega$ and $\exp(iJ)$ in IIA and vice-versa in IIB.
Then, defining $\d_H\equiv \d+H\wedge$,  the background supersymmetry conditions schematically read (see eq.~\eqref{susyconds} below for the
 precise expressions)
\bea\label{introeq}
\d_H\Psi_1&\simeq& (1/R)\Re\Psi_2+\ \text{RR-fluxes}\ ,\cr
\quad \d_H\Psi_2&\simeq& (1/R)\Im\Psi_1\ ,
\eea
where $R$ is the AdS-radius. If a pure spinor is $\d_H$-closed, then it defines an integrable GC structure\footnote{In fact,
the pure spinor being closed is slightly stronger so the converse is not true. For the precise statement see \cite{hitchin,gualtieri}.}.
Thus, in the flat-space limit ($R\rightarrow \infty$) $\Psi_2$ defines and integrable GC structure,
while the Ramond-Ramond (RR) fluxes create an obstruction to the integrability of the GC structure associated to $\Psi_1$.
On the other hand, for $R$ finite there are extra `geometric fluxes' on the right-hand side of (\ref{introeq}) acting
as obstructions to the integrability of both GC structures. This will complicate even further the general study of the moduli (along the lines of
e.g.~\cite{tomasiello}).

In this paper we will study D-branes on this general class of AdS flux compactifications.  Supersymmetric D-branes should minimize their energy inside their deformation class and this physical principle is usually realized by the existence of (generalized) calibration forms \cite{harvey,papa,gencal,luca1} which must obey appropriate differential conditions.
This is expected to happen quite generically and it was indeed shown in \cite{luca1} that in the Minkowski limit of the above class of vacua each of the background supersymmetry equations provides exactly the right differential condition for having a proper calibration for a different type of D-brane as viewed from the four-dimensional space-time: space-filling, domain wall or string-like. In particular, the presence of the RR-fluxes in (\ref{introeq}) is required because of the Chern-Simons term in the D-brane action, providing an extension of the general idea presented in \cite{papa}.

However, when $R$ is finite the appearance of the additional `geometric fluxes' on the right-hand side of (\ref{introeq}) seems puzzling in view of this relation
between background supersymmetry and D-brane calibrations.
In this paper we discuss how these additional terms are related to one of the deep differences between AdS and flat space-time: that
AdS has a boundary that can be reached in finite time. This changes considerably the energy and stability properties in comparison to the flat case
and was studied some time ago in a series of papers \cite{bf,freedman} both at the perturbative and non-perturbative level, in supergravity and rigid theories.
For example, it was shown that the mass of the fields and more generally the potential density of a theory in AdS need not be non-negative
in order to have a stable vacuum.
As we will show, these four-dimensional features have an elegant realization in our D-brane setting:
just as in the flat case we will see that the eqs.~(\ref{introeq}) can be equivalently seen as integrability conditions
for a set of generalized calibrations that provide a natural setting for studying supersymmetry, energetics and stability of D-branes in AdS flux compactifications.
Even though we consider here explicitly the case of compactifications to AdS$_4$,
the physical arguments are quite general. Indeed, as we prove in appendix \ref{susycalgen},
any time-independent Killing spinor generating a supersymmetry of a static background geometry can be used to
construct a corresponding calibration characterizing the static D-brane configurations preserving it.

The modification of the energetics and stability of D-branes essentially induced by the boundary of AdS
can have profound implications on the topology of stable supersymmetric D-branes in the internal space.
Indeed, consider for concreteness space-filling branes in compactifications to flat space.
To be stable they must usually wrap an internal non-trivial cycle unless background fluxes enter the game through
the dielectric Myers effect \cite{myers}. However, in compactifications to AdS space
it often happens that they wrap internal trivial cycles. For example, we will consider somewhat more explicitly  $SU(3)$-structure IIA flux compactifications.
They include Freund-Rubin-like compactifications on $S^6$ which admit supersymmetric D-branes wrapping a trivial $S^3\subset S^6$ (see also \cite{acharyadenef}).

Let us now sketch in more detail the concrete mechanism. Unlike in Minkowski space a D-brane extending to infinity
in its external AdS part has a boundary at the boundary of AdS, even though in the internal part it wraps a cycle.
In itself the D-brane is not consistent because it violates invariance under RR gauge transformations on this boundary.
Intuitively RR-charge and by supersymmetry also energy can leak away.
To have a consistent description of the dynamics in cases like this, the standard procedure is to add boundary conditions.
In our setting this implies that when considering fluctuations one would not be allowed to change the D-brane internal embedding at the boundary.
As a result, these D-brane fluctuations have non-zero gradient energy since they have to have some profile
in order to vanish at the boundary, and while naively they might seem tachyonic they are lifted to be massless or even massive when
this gradient energy is taken into account. This is in fact the Breitenlohner-Freedman \cite{bf} mechanism.

A perhaps more physical intuition about how this mechanism works can be obtained
by generalizing the above prescription in such a way as to allow fluctuations that are {\em not} vanishing at the boundary.
An example would be the fluctuations that one might naively consider, the ones constant over the AdS under which the D-brane moves homogeneously in the internal space.
The price to pay is that the system must be completed by a D-brane lying at the boundary of AdS that acts
as a sink for the charge and the energy. This D-brane thus restores the RR gauge invariance discussed above.
Now, if one looks more closely at the bulk supersymmetry conditions of the form (\ref{introeq})
it turns out that they are naturally associated to the calibration for the network of both the original D-brane extending to infinity and the additional D-brane at the boundary.
These networks can be studied in generalized geometry along the lines of \cite{jarahluc}.

{}From both points of view one has to abandon (at least for the fluctuations) the picture that is used in the Minkowski analysis,
where the D-branes are constant over the space-time part.
It is thus natural to study generalized calibrations in a 1+9-dimensional setting, which, taking into account
that one needs a time-like Killing vector for studying static calibrations, is the most general.
Extending the relation between the supersymmetry conditions for the bulk and the differential calibration conditions
to this general setting is exactly the content of appendix \ref{susycalgen}.

{}Another peculiar effect of AdS is that in the effective four-dimensional supergravity theory,
the D-flatness condition for supersymmetric AdS vacua is automatically implied by the F-flatness.
In \cite{effective} we showed that for the closed string sector this result can be uplifted to ten dimensions.
Extending the identifications of D- and F-terms for D-branes found in \cite{luca2} to the case of  AdS$_4$ flux vacua,
we find in this paper that also for the open string sector the D-flatness condition is implied by the F-flatness condition.
As we will see another related observation is that in the AdS case there are no string-like supersymmetric D-branes.
These results are of course related to the additional terms proportional to $1/R$, which appear on the right-hand side of (\ref{introeq}),
giving another nice example of the deep relation between the closed and open string sector.

We provide examples of calibrated D-branes in a special class of type IIA $SU(3)$-structure backgrounds, the nearly K\"ahler geometries.
Many of these examples are similar to much better
studied supersymmetric D-brane configurations in AdS$_5$ backgrounds, see e.g.~\cite{kr,mtaylor,urangacasc}, and in fact nearly
K\"ahler geometry can be seen as the six-dimensional analogue of Sasaki-Einstein geometry.

\section{The background AdS$_4$ flux vacua}
\label{back}

We start by describing the background geometry, adopting the language of generalized complex geometry as in \cite{gmpt}.
This already proved to be a very natural framework for compactifications to Minkowski $\mathbb{R}^{1,3}$
especially once D-branes are taken into account \cite{gencal,luca1}.
Indeed, in that case each of the background supersymmetry equations corresponds to a type of calibrated D-brane \cite{luca1,luca2}.
As we will show in section \ref{adstot}, also for AdS compactifications a natural interpretation can be found for the extra terms and the relation still holds.

We consider type II theories on ten-dimensional space-times of (warped) factorized form AdS$_4\times_w M$, where $M$ is the internal six-dimensional space. Thus the ten-dimensional metric has the form
\bea
\d s^2_{(10)}=e^{2A(y)}\d s^2_{(4)}+g_{mn}(y)\d y^m\d y^n\ ,
\eea
where $\d s^2_{(4)}$ is the AdS$_4$ metric.
All the background fluxes preserve the conformal symmetry $SO(2,3)$ of AdS$_4$ and depend only on the internal coordinates $y^m$. Thus the NSNS $H$-field has only internal indices and the RR fields, which in the democratic formalism of \cite{democratic} can be
conveniently organized in the polyform $F=\sum_{n}F_{(n)}$ ($n$ is even in IIA and odd in IIB),  split as follows
\bea
F=\text{vol}_4\wedge e^{4A}\tilde F+\hat F\ .
\eea
Here $\text{vol}_4$ is the (unwarped) AdS$_4$ volume form and $\tilde F$ and $\hat F$ have only internal indices.
In the democratic formalism the RR-fields are doubled and this is
compensated by imposing a Hodge duality condition, which for the internal and external components of the RR fields implies
$\tilde F=\mp\sigma(\star_6 \hat F)$ in IIA/IIB\footnote{Here and in the following, the upper sign is for IIA while the lower is for IIB.}, where $\sigma$ is the operator acting on forms by reversing the order of their indices.

Since we require minimal $\caln=1$ four-dimensional supersymmetry, our background admits four independent Killing spinors of the form
\bea\label{adskilling}
\epsilon_1 &=&\zeta_+\otimes \eta^{(1)}_+ +\, (\text{c.c.})\ ,\cr
\epsilon_2 &=&\zeta_+\otimes \eta^{(2)}_\mp+\, (\text{c.c.})\ ,
\eea
for IIA/IIB. In the above the two internal chiral spinors $\eta^{(1)}_+$ and $\eta^{(2)}_+$ are fixed for a certain background geometry.
They define a reduction of the structure group of $T_M\oplus T_M^\star$ from $SO(6,6)$ to $SU(3)\times SU(3)$ \cite{gualtieri} and thus characterize the solution.
$\zeta_+$ on the other hand is any of the four independent AdS$_4$ Killing spinors satisfying the equation
\begin{equation}
\nabla_\mu \zeta_-=\pm \frac12 W_0 \gamma_\mu \zeta_+ \label{spinori}\ ,
\end{equation}
for IIA/IIB.\footnote{We indicate the four-dimensional (unwarped) curved gamma-matrices with $\gamma_\mu$ and the six-dimensional ones with $\hat{\gamma}_i$.
See \cite{luca1} for detailed conventions.} $W_0$ is proportional to the on-shell value of the superpotential $\calw$ in the four-dimensional description of \cite{effective}
so that $|W_0|^2 = - \Lambda/3$ with $\Lambda$ the effective four-dimensional cosmological constant.

In \cite{gmpt} the supersymmetry equations obtained from putting the supersymmetry variations of the fermions to zero were written in terms of the $SO(6,6)$ pure spinors $\Psi^\pm$.
These $SO(6,6)$ spinors can also be seen as polyforms and related to those by the Clifford map as the internal $SO(6)$ bispinors
$\eta^{(1)}_+\otimes\eta^{(2)\dagger}_\pm$. In the case of AdS$_4$  the two internal spinors {\em must} have the same norm \cite{granascan}: $\eta^{(1)\dagger}_+\eta^{(1)}_+=\eta^{(2)\dagger}_+\eta^{(2)}_+=|a|^2$, while for compactifications to Minkowski
this is imposed as
an additional requirement necessary for the background to admit static supersymmetric D-branes \cite{luca1}.
It is convenient to introduce the normalized pure spinors\footnote{Note that in \cite{jarahluc,luca2,deforms,lucanapoli} the normalized pure spinors (\ref{nps}) were denoted by $\hat\Psi^{\pm}$,  while $\Psi^{\pm}$ referred to the ones via the Clifford map associated to $\eta^{(1)}_+\otimes\eta^{(2)\dagger}_\pm$ without normalization.}
\bea
\label{nps}
\slashchar{\Psi}^{\pm}=-\frac{8i}{|a|^2} \; \eta^{(1)}_+\otimes\eta^{(2)\dagger}_\pm \ ,
\eea
and rename them as
\bea
\Psi_1=\Psi^{\mp}\quad\text{and}\quad \Psi_2=\Psi^{\pm}\qquad\text{in IIA/IIB.}
\eea
The supersymmetry conditions found in \cite{gmpt} can be rewritten as the following minimal set of equations
\begin{subequations}
\label{susyconds}
\begin{eqnarray}
\label{susycond0}
\d_H \big(e^{4A-\Phi} \Re \Psi_1 \big) & =& (3/ R) \, e^{3A-\Phi} \Re (e^{i\theta} \Psi_2) + e^{4A} \tilde{F} \, ,\\
\d_H \big[e^{3A-\Phi}\Im (e^{i\theta}\Psi_2)\big]&=& (2/ R) \,e^{2A-\Phi}\Im \Psi_1\ , \label{susycond1}
\end{eqnarray}
\end{subequations}
where $\d_H=\d+H\wedge$ and we put $W_0=\frac{e^{-i\theta}}{R}$ with $R$ the AdS radius. They imply as integrability conditions\footnote{We take into account the equations of motion for $\hat{F}$.}
the two further equations
\bea\label{intcond}
\d_H(e^{2A-\Phi}\Im \Psi_1)=0 \, , \qquad \d_H\big[e^{3A-\Phi} \Re (e^{i\theta} \Psi_2)\big]=0\ .
\eea
So in the AdS case we have two minimal real equations for the pure spinors, while the two additional equations~(\ref{intcond}) come as a necessary requirement.
This is significantly different from the Minkowski case, which we can find by taking the $R\rightarrow \infty$ limit, where one has four independent
polyform equations.
As explained in \cite{effective} the first equation of \eqref{intcond} can be interpreted as a D-flatness condition, while the other equations in \eqref{susyconds} and \eqref{intcond} are F-flatness conditions.
In general, the D-flatness condition indeed follows from the F-flatness conditions if $\calw \neq 0$ as is the case for AdS compactifications.

As a guiding example it can be useful to keep in mind the form of the above pure spinors in the $SU(3)$-structure case, in which the internal spinors are parallel. Putting $\eta^{(1)}_+=a\eta_+$ and $\eta^{(2)}_+=b\eta_+$ with $|a|=|b|$ and $\eta_+^\dagger\eta_+=1$, and further defining
the (almost) symplectic two-form $J_{mn}=i\eta^\dagger_+\hat\gamma_{mn}\eta_+$ and $(3,0)$-form $\Omega_{mnp}=i\eta_-^{\dagger}\hat\gamma_{mnp}\eta_+$
characterizing the $SU(3)$-structure, one has
\bea
\Psi^+=-i(a/b)e^{iJ}\, , \qquad \Psi^-=(a/b^*)\Omega\ .
\eea

Such a restriction to parallel internal spinors however, while allowing AdS vacua in IIA \cite{cvetic,tsimpis}, automatically excludes AdS vacua in IIB as can be easily seen from (\ref{susycond1}), at least if such a structure is not deformed by non-perturbative corrections \cite{effective}. To have IIB classical AdS vacua one is then forced to consider a more general $SU(2)$- or $SU(3)\times SU(3)$-structure background.

The extra terms on the right-hand side of \eqref{susyconds} will force us to rethink the relation between bulk supersymmetry conditions and calibration conditions.
So first let us introduce calibrated D-branes as in \cite{gencal,luca1}.

\section{Introducing supersymmetric D-branes}
\label{susyd}

Let us now introduce static supersymmetric probe D-branes on the backgrounds described in section \ref{back}
following the procedure based on $\kappa$-symmetry of \cite{gencal,luca1}.
Since the details turn out to be almost identical, we will omit them here, referring to those papers or to appendix \ref{reviewcal} where
we review and extend the procedure to more general backgrounds.

Disallowing world-volume field-strength along AdS$_4$ we see from the usual $\kappa$-symmetry argument that, locally, the supersymmetry conditions
for the D-branes look exactly the same as the ones  for flat space \cite{luca1}.
Thus we know that we can only have supersymmetric D-branes which from the four-dimensional point of view
are  either space-filling D-branes, strings or domain walls. The actual shape of the last case in the four dimensions
will be considered in section \ref{poincarecord}.

Focusing on the internal manifold $M$,  an  internal generalized $p$-dimensional cycle\footnote{We use the terminology of \cite{gualtieri}
where a generalized submanifold $(\Sigma,\calf)$ is nothing but a submanifold $\Sigma$ with a world-volume field-strength $\calf$
satisfying the modified Bianchi identity $\d\calf=H|_\Sigma$, which can be extended to include monopole sources if one considers networks of D-branes \cite{jarahluc}.} $(\Sigma,\calf)$ wrapped by a supersymmetric D-brane must satisfy a calibration-like condition of the form
\bea\label{intcalcond}
\big[ \omega|_{\Sigma}\wedge e^\calf\big]_{\text{top}}=\cale_{\text{DBI}}(\Sigma,\calf) \, ,
\eea
where\footnote{For simplicity we put all the D-brane tensions to one. The correct tensionful prefactors can be easily reintroduced.}
\bea
\cale_{\text{DBI}}(\Sigma,\calf)=e^{qA-\Phi}\sqrt{\det(g|_\Sigma+\calf)}\,\d^p\!\sigma\ ,
\eea
 with $q=4,3,2$ for space-filling, domain wall and string-like D-brane configurations respectively, while correspondingly
\bea\label{6dcal}
\omega^{\text{(sf)}}=e^{4A-\Phi}\Re \Psi_1\, , \qquad  \omega_\varphi^{\text{(DW)}}=e^{3A-\Phi}\Re(e^{i\varphi} \Psi_2)\, , \qquad \omega^{\text{(string)}}=e^{2A-\Phi}\Im \Psi_1\,  ,
\eea
where $\varphi$ is a constant phase.

{}From (\ref{intcalcond}) one gets that the four-dimensional effective tension is given by
\bea
T_{\text{4d}}=\int_{\Sigma} \omega|_\Sigma\wedge e^\calf\ ,
\eea
and should not vanish for a non-degenerate physically meaningful configuration.
However, from (\ref{susycond1}) one immediately sees that $\omega^{\text{(string)}}$ is $\d_H$-exact
and thus $T_{\text{4d}}^{\text{(string)}}=0$.
We conclude that supersymmetric tension-full strings {\em cannot} be obtained by wrapping D-branes on internal cycles,
and this suggests that any non-BPS D-brane string cannot be stable and will eventually annihilate {\em locally} on its world-sheet.
Thus, while in Minkowski compactifications stable BPS D-brane strings are naturally admitted, in the AdS$_4$ case they are not.
This is an example of one of several deep differences between Minkowski and AdS$_4$ flux compactifications and has a counterpart in the expected four-dimensional description of the system. We will come back to this in the next section. For the same reason $\varphi=\theta$ tension-full domain walls can be obtained only if $\tilde F$ is not $\d_H$-exact.

So far we have only studied the purely algebraic aspect of the supersymmetry condition. But the calibration-like condition (\ref{intcalcond}) requires some more discussion to be interpreted as a proper calibration condition. Indeed, we still have to show the stability of the calibrated configurations under continuous deformations. As explained in appendix \ref{reviewcal},
if we tried to insist on a local four-dimensional picture where the main properties can be obtained looking only at the internal six-dimensional part,  this would require the calibration forms $\hat{\omega}$, which are analogously to \eqref{calform} defined as
\eq{
\hat{\omega}^{(\text{sf})} = \omega^{\text{(sf)}} - e^{4A} \tilde{C} \, , \qquad \hat{\omega}^{(\text{DW})}_{\varphi} = \omega^{\text{(DW)}}_\varphi \, ,
}
to be $\d_H$-closed. This is indeed the case for Minkowski compactifications, but not for AdS because of the appearance of `geometric fluxes'
on the right-hand side of (\ref{intcond}) whose role has to be properly clarified to have a consistent global picture.
We will address this problem in section \ref{adstot}
where we will see that those geometric fluxes have a very natural interpretation in a full ten-dimensional picture. But let us first take a closer look at the four-dimensional interpretation of the present results.

\section{Four-dimensional interpretation: F- and D-terms}
\label{4d}

In this section we write the supersymmetry/calibration conditions for D-branes filling AdS$_4$ as F- and D-flatness conditions.
We will show  that because of the AdS$_4$ geometry the D-flatness condition for the open string modes follows from the F-flatness conditions,
as expected on general grounds and as was also shown for the closed string moduli in \cite{effective}.

The calibration conditions obtained from (\ref{intcalcond}) by using the three calibrations (\ref{6dcal}) admit an interesting alternative formulation \cite{luca1,luca2}.
Indeed, an equation of the form
\bea\label{cc1}
[\Re(e^{i\alpha}\Psi^\pm)|_\Sigma\wedge e^\calf]_{\text{top}}=\sqrt{\det(g|_\Sigma+\calf)}\ ,
\eea
for some constant phase $e^{i\alpha}$, can be rewritten as the following pair of conditions (supplemented with a condition on the orientation)
\bea\label{altersusy}
[(\mathbb{X}\cdot\Psi^\mp)|_\Sigma\wedge e^\calf]_{\text{top}}=0\, , \qquad [\Im(e^{i\alpha}\Psi^\pm)|_\Sigma\wedge e^\calf]_{\text{top}}=0\ ,
\eea
for any $\mathbb{X}=X+\xi\in T_M\oplus T^\star_M$, with $\mathbb{X}\cdot=\iota_X+\xi\wedge$ .

For space-filling branes an effective $\caln=1$ four-dimensional description should exist. The complete system
should in principle include both open and closed string modes, although for the moment we focus on the open string modes and freeze the closed string modes.
The effective theory is described by the K\"ahler potential $\calk$, the superpotential $\calw$ for the chiral multiplet sector, the D-term $\cald$ and the holomorphic metric $f$ for the vector multiplet.  If $\calg$ indicates the metric for the chiral multiplets, the effective potential takes the form
\bea\label{effpot}
V=e^\calk\big[\calg^{-1}(D\calw,D\bar\calw)-3|\calw|^2\big]+\frac12 (\Re f)^{-1}(\cald,\cald)\ .
\eea
\cite{luca2} compared the potential (\ref{effpot}) with the potential found from the D-brane action
and also studied the supersymmetry transformation rules,
finding in both cases the following identifications for the D- and F-terms
\bea\label{dfterms}
\cald=[(e^{2A-\Phi}\Im\Psi_1)|_\Sigma\wedge e^\calf]_{\text{top}}\, , \qquad
e^{\calk/2}D\calw (\mathbb{X})=[(e^{3A-\Phi}\mathbb{X}\cdot\Psi_2)|_\Sigma\wedge e^\calf]_{\text{top}}\ ,
\eea
where the F-terms are defined by considering the covariant derivative $D$ along an arbitrary section $\mathbb{X}$ of the generalized normal bundle $\caln_{(\Sigma,\calf)}$,
which contains the infinitesimal deformations of the generalized cycle $(\Sigma,\calf)$ (see \cite{luca2,deforms} for the definition and more details). Note that the discussion of \cite{luca2} about the holomorphy of the superpotential is still valid and one can locally define a (in general non-integrable) complex structure on $\caln_{(\Sigma,\calf)}$, such that the F-term in (\ref{dfterms}) only contains holomorphic indices.

Until now we have only proved the {\em algebraic} equivalence (up to orientation choice) of the supersymmetry/calibration condition (\ref{intcalcond}) for space-filling D-branes and the pair of conditions
\begin{subequations}
\begin{align}
\cald & = 0 & & (\text{D-flatness})\ , & & \\
e^{\calk/2}D\calw (\mathbb{X}) & = 0 \, , \qquad \forall \mathbb{X}\in T_M\oplus T^\star_M & & (\text{F-flatness})\ . & &
\end{align}
\end{subequations}
However, taking into account also the {\em differential} background supersymmetry equations more can be said. Indeed, let us choose $\mathbb{X}_\lambda=\d\lambda$ for some $\lambda$. Then, using (\ref{susycond1}) we have that
\bea\label{fimpliesd}
\int_\Sigma e^{\calk/2}D\calw (\mathbb{X}_\lambda)&=& -2i \frac{e^{-i \theta}}{R} \int_\Sigma \lambda \cald\ .
\eea
Since $\lambda$ can be an arbitrary function we arrive at the conclusion that for AdS flux compactifications (for which $R$ is finite)
{\em F-flatness implies D-flatness}! Thus, in order to check that a space-filling D-brane is supersymmetric (up to the appropriate choice of orientation) it is sufficient to check the F-flatness condition.

This remarkable result is not so unexpected if one considers the problem from the four-dimensional point of view.
Indeed, even though we are considering a probe D-brane, we expect it to be described by a complete supergravity admitting an AdS vacuum.\footnote{The closed string may be decoupled by sending both the cosmological constant and the Planck mass to infinity in order to be left with a rigid theory on an AdS background.}
In a general $\caln=1$ supergravity there is a relation between F- and D-terms which has {\em exactly} the form (\ref{fimpliesd})
(see e.g.~\cite{dealwis} where this point is particularly stressed).
The identification can be made precise using the results of \cite{luca2} which allow to properly identify the D-term as the moment map
associated to the world-volume gauge transformations, parameterized by $\lambda$ in (\ref{fimpliesd}). Even though \cite{luca2}
focused on $\mathbb{R}^{1,3}$ flux compactifications, the analysis is still valid here.

This also explains from a four-dimensional point of view why we cannot have stable BPS D-brane strings in AdS flux compactifications.
Indeed, they should correspond to D-term solitonic strings that must be F-flat everywhere \cite{dstring} and whose tension is essentially given by a Fayet-Iliopoulos term. Since F-flatness implies D-flatness, such a non-vanishing Fayet-Iliopoulos term is not possible.

\section{A closer look at D-brane stability}
\label{adstot}

\subsection{D-brane energy problem and sketch of its solution}
\label{sketchsol}

In sections \ref{susyd} and \ref{4d} we have discussed the supersymmetry conditions that must be imposed on the internal generalized cycles wrapped by D-branes that are space-filling, domain wall or string-like in AdS$_4$ -- concluding that the latter are dynamically excluded -- from a local and algebraic point of view. Looking at the differential conditions however, we found that at first sight the would-be generalized calibration form is not $\d_H$-closed so that the calibrated configurations are
not at a minimum of the energy. In this subsection we state the problem and sketch the main principle of the solution, for which the global structure of AdS$_4$ turns out to be important. The details are worked out in the next subsections.

We consider for definiteness space-filling D-branes, while the analysis for domain wall D-branes is similar.
It is tempting to associate to a configuration wrapping the internal generalized cycle $(\Sigma,\calf)$ an effective four-dimensional
potential
\bea\label{4dpot}
\calv(\Sigma,\calf)=\int_{\Sigma}  e^{4A-\Phi}\sqrt{\det(g|_\Sigma+\calf)}-\int_\Sigma e^{4A}\tilde C|_\Sigma\wedge e^\calf\ ,
\eea
where $\tilde C$ is the (locally defined) RR-potential such that $\tilde F=e^{-4A}\d_H(e^{4A}\tilde C)$.
However, such a potential turns out {\em not} to be naturally bounded from below by its value on supersymmetric configurations satisfying (\ref{intcalcond}). The reason is that $\hat{\omega}^{\text{(sf)}}=\omega^{\text{(sf)}}-e^{4A}\tilde C$ is not $\d_H$-closed due to the term proportional to $\Re(e^{i\theta}\Psi_2)$ on the right-hand side of (\ref{susycond0}). Thus  $\hat{\omega}^{\text{(sf)}}$ cannot be seen as a proper generalized calibration and we cannot use the usual argument showing that calibrated configurations are energy density minimizing in their generalized homology class as in \cite{gencal,luca1}. Similarly for domain walls the problem comes
from the term on the right-hand side of (\ref{susycond1}).

Although the calibrated configuration is not in general at a minimum of the potential, one can show that it is still at least
at a stationary point so that it is a solution of the equations of motion.
Indeed, if we vary along a section $\mathbb{X}$ of the generalized normal bundle $\caln_{(\Sigma,\calf)}$, which as
shown in \cite{luca2} describes the general deformation of $(\Sigma,\calf)$, we find
\eq{
\label{soleom}
\delta_{\mathbb{X}} \calv(\Sigma,\calf) = \int_\Sigma \call_{\mathbb{X}} \hat{\omega}^{(\text{sf})} \wedge e^\calf
= \int_\Sigma \mathbb{X} \cdot \d_H \hat{\omega}^{(\text{sf})} \wedge e^\calf = \frac{3}{R} \int_\Sigma \Re (e^{i \theta} e^{\calk/2} D \calw(\mathbb{X})) = 0 \, ,
}
where the last expression is zero because of the F-flatness. However, around a stationary point that is not a minimum there are tachyonic modes. It is well known however
that in field theories on AdS space tachyonic modes do not signal instability as they would in Minkowski space as long as they are above the
Breitenlohner-Freedman bound \cite{bf}. In a moment we will argue that indeed they are.

In the following we will indicate the submanifold that the D-brane wraps in the AdS part by $\Pi$.
The solution to the problem comes naturally if we consider the global structure of AdS$_4$, which can be considered as a cylinder
having a boundary at infinity with topology $\mathbb{R}\times S^2$. Thus a space-filling or a domain wall D-brane extending to infinity that is homogeneous -- which means wrapping the same internal generalized cycle $(\Sigma,\calf)$ in all its space-time points and having vanishing world-volume gauge field along AdS$_4$ -- has a boundary at the boundary of AdS$_4$: $\partial (\Pi \times \Sigma) = \partial \Pi \times \Sigma=(\Pi \cap \partial \text{AdS}_4) \times \Sigma$. Therefore, on its own it is not a cycle, which leads to a problem of invariance under RR gauge transformations (for a discussion in the generalized context see \cite{jarahluc}). Indeed, if we consider the variation under $\delta C = \d_H \lambda$ of the Chern-Simons term in the action of such a D-brane we find
\eq{
\label{RRtrans}
\delta_\lambda S_{\text{CS}} = \int_{\partial \Pi \times \Sigma} \lambda|_{\partial \Pi \times \Sigma} \wedge e^\calf \neq 0 \, .
}
This breaking of the gauge invariance implies a breaking of charge conservation and by supersymmetry also a leaking of energy at the
boundary of AdS$_4$.

There are basically two ways out. One is to consider only gauge transformations and for consistency also fluctuations that vanish at the
boundary. This amounts to imposing fixed boundary conditions. The other is to introduce a domain wall respectively string-like D-brane
at the boundary of AdS$_4$
that acts as a sink for RR-charge and energy. Both of these require to extend the class of allowed deformations to the non-homogeneous ones,
even though the original configuration we expand around is homogeneous. So we must consider the full nine-dimensional space (or at least the seven-dimensional space including the internal space and the radial coordinate) as a whole. How to construct a total calibration
on such a space is explained in detail in appendix \ref{susycalgen}. For the case of AdS$_4$ compactifications the calibration
is given by \eqref{totcalp} and \eqref{totcalglob} below.

\begin{figure}
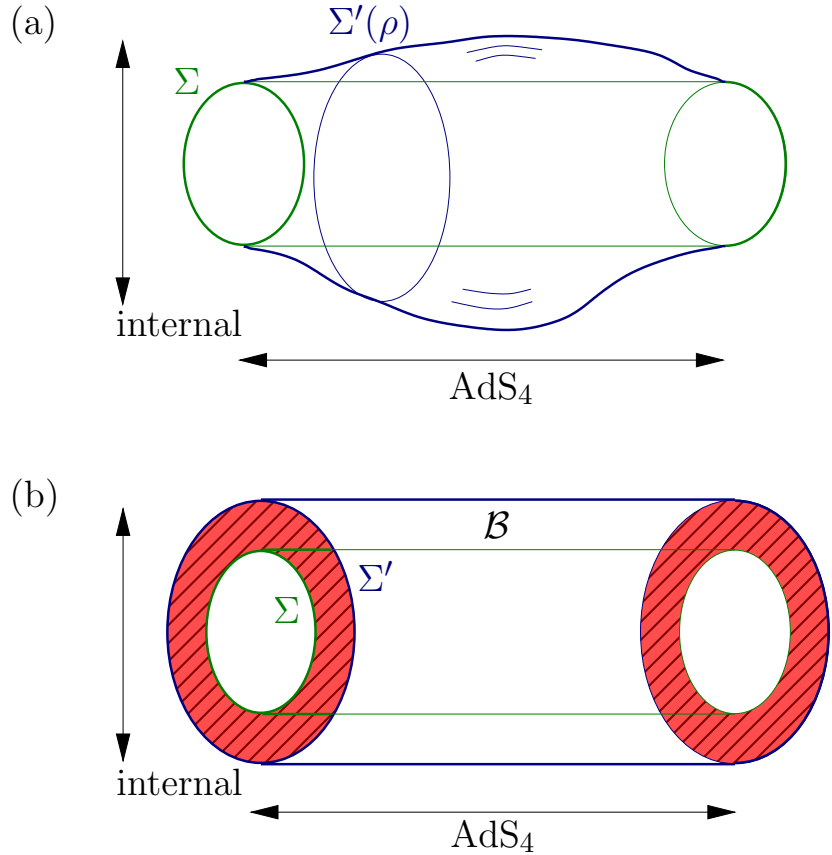

\setlength{\fboxsep}{14pt}%
\ovalbox{
\begin{minipage}{0.93\textwidth}
\begin{center}
\psset{unit=0.7cm}
\include{boundaryfluccolor}
\include{homfluccolor}
\caption{In both pictures the external AdS$_4$ is represented as a line segment, while
the internal cycles are represented as circles.
(a) Fluctuation that vanishes at the boundary.  The shape of the deformed cycle $\Sigma'(\rho)$ depends on the
location in AdS$_4$. Since the fluctuation is required to vanish at the boundary of AdS$_4$, the deformed
cycle has to coincide there with the original cycle. (b) Homogeneous fluctuation. $\Sigma$ is homogeneously deformed into $\Sigma^\prime$, which means
in particular that the deformation does not vanish at the boundary of AdS$_4$. Suppose $\calb$ is the space
between $\text{AdS}_4 \times \Sigma$ and $\text{AdS}_4 \times \Sigma^\prime$. {}From the picture it is clear that the boundary
of $\calb$ is not only the difference between the original D-brane and its deformation, but also includes a difference of
boundary D-branes (shaded area).}
\label{fluctuations}
\end{center}
\end{minipage}}
\end{figure}
{}From the point of view of the low-energy field theory in AdS$_4$ the picture where we impose fixed boundary conditions
on the fluctuations is the most natural one. In this case the deformed internal cycle $\Sigma'(\rho)$ depends on the radial coordinate $\rho$
of AdS$_4$. See figure \ref{fluctuations}(a). As we will discuss explicitly in the next subsections, this non-trivial radial dependence
gives an extra contribution to the total energy density, make the latter positive definite.
One can also look in the way of Breitenlohner and Freedman \cite{bf}:
fluctuation modes that naively look tachyonic acquire extra gradient energy because of the non-trivial profile $\Sigma'(\rho)$ and
are lifted to massless or even massive modes. These modes are said to obey the Breitenlohner-Freedman bound.

Next, suppose that we want to consider fluctuations that do {\em not} vanish at the boundary, an example would be the homogeneous fluctuations
we originally considered. See figure \ref{fluctuations}(b). Since the difference of the submanifolds wrapped by the deformed and the original D-brane $\Pi \times(\Sigma' - \Sigma)$ has a boundary, it can never be a boundary itself, invalidating the usual calibration argument \eqref{calboundshow}.
The way out is to add to the original D-brane a boundary D-brane wrapping $\pm \partial \Pi \times \Gamma$ (upper/lower sign for space-filling/domain wall D-brane) with  $\Gamma$ a submanifold in the internal space so that $\partial \Gamma=\Sigma$. This also solves the RR gauge transformation problem since $\partial(\pm \partial \Pi \times \Gamma)= - \partial \Pi \times \Sigma$ so that the transformation of the boundary D-brane exactly
compensates the one of the original D-brane \eqref{RRtrans}.
Such a boundary D-brane reminds us of the `membrane at the end of the universe' studied in the eighties (see e.g.~\cite{duff} for a review and references therein for the original work). One implication is that we should also take the energy density of the boundary D-brane into account so that the naive four-dimensional effective potential of eq.~\eqref{4dpot} is modified into
\eq{\label{4dpot2}
\tilde\calv(\Sigma,\calf)=\int_{\Sigma}  e^{4A-\Phi}\sqrt{\det(g|_\Sigma+\calf)}-\int_\Sigma \cala|_\Sigma\wedge e^\calf\ ,
}
where the `modified' RR-potential $\cala$ is defined such that
\bea\label{modpot}
\d_H\cala=\frac{3}{R}e^{3A-\Phi}\Re(e^{i\theta}\Psi_2)+e^{4A}\tilde F\ .
\eea
We will see below that the part of $\cala$ corresponding to the first term on the right-hand side of \eqref{modpot} is indeed associated
to a boundary D-brane.  Moreover, we will see that the modified potential is positive definite and bounded from below by its value for calibrated configurations.

Field theories on AdS space-times have been studied in \cite{freedman}, where the same problem of finding a manifestly positive definite potential density was addressed.
It was shown that the proper energy density, obeying the proper commutation relations with the other charges of the theory, contains a counterterm which corrects the `naive' potential density to a positive definite quantity.
The above discussion thus provides a nice geometrical interpretation of this counterterm. Indeed, the corrective term in (\ref{modpot}) has exactly the form  $(3/R)\Re(e^{i\theta}\calw)$ that was found for the corrective term in the Wess-Zumino model studied in \cite{freedman}.

\subsection{Total calibration in Poincar\'e coordinates}
\label{poincarecord}

In the previous subsection we saw that the solution of the stability problem required considering generalized calibrations in the complete
nine-dimensional space (or at least the seven-dimensional space consisting of the radial coordinate and the internal space). Such a complete generalized calibration can be constructed by following the general procedure described in appendix \ref{susycalgen}, which we will apply here to compactifications on AdS$_4$. In this subsection we use Poincar\'e coordinates to describe AdS$_4$, avoiding some technical subtleties that arise when adopting global coordinates. However, most of the results of the present subsection can be applied to AdS$_4$ in global coordinates too as we will show in the next subsection.

The metric in Poincar\'e coordinates  $(t,x^1,x^2,z)$ can be written as
\bea
\d s^2_{(4)}=e^{\frac{2z}{R}}(-\d t^2+\d \vec x^2)+\d z^2\ ,
\eea
where all the coordinates extend to all of $\mathbb{R}$.
To construct the calibration we need the explicit form of the four-dimensional spinors $\zeta_+$ satisfying the Killing spinor equation~(\ref{spinori}),
which can be found for example in \cite{town}. Two are time-independent and we can use them to directly construct two corresponding total generalized calibrations, which are related to each other by a rotation
in the $(x^1,x^2)$-plane. Thus, any calibration can be obtained by applying a rotation to the following reference calibration:
\bea\label{totcalp}
\Theta_{P}=\Theta_{P}^{\text{(sf)}}+\Theta_{P}^{\text{(DW)}}\ ,
\eea
with
\bea
\label{totcalpsf}
\Theta_{P}^{\text{(sf)}}=e^{\frac{3z}{R}}\d x^1\wedge \d x^2\wedge\d z\wedge\omega^{\text{(sf)}} +e^{\frac{3z}{R}}\d x^1\wedge\d x^2\wedge \omega^{\text{(DW)}}_{\varphi=\theta}\ ,
\eea
and
\bea
\label{totcalpdw}
\Theta_{P}^{\text{(DW)}}=e^{\frac{2z}{R}}\d x^1\wedge \d z\wedge\omega_{\varphi=\theta-\pi/2}^{\text{(DW)}} +e^{\frac{2z}{R}}\d x^1\wedge \omega^{\text{(string)}}\ ,
\eea
where the different $\omega$ are given in \eqref{6dcal}. Note that the $SO(2)$ invariance is explicitly broken only by $\Theta_P^{\text{(DW)}}$.

A $\kappa$-symmetry argument analogous to the one used for six internal dimensions in \cite{gencal,luca1}
can be used to show that D-branes calibrated with respect to $\Theta_P$ are supersymmetric. Furthermore using (\ref{susycond0}) one can easily check that  $\d_H\Theta_P^{\text{(sf)}}=-e^{\frac{3z}{R}}\d x^1\wedge \d x^2\wedge\d z\wedge e^{4A}\tilde F$ and $\d_H\Theta_P^{\text{(DW)}}=0$. Thus  $\Theta_P$ is a well-defined ``total'' generalized calibration as in appendix \ref{reviewcal}, so that calibrated D-brane networks will be at a minimum of the energy and thus stable.
D-branes with fixed boundary conditions should thus calibrate \eqref{totcalp}, which sees the non-trivial radial dependence and thus properly picks up the gradient energy.

Note also that the above differential conditions on $\Theta_P$ are {\em equivalent} to the requirement that the background supersymmetry conditions (\ref{susyconds}) are satisfied. Thus we find that, like in the $\mathbb{R}^{1,3}$ case, the AdS flux compactifications are {\em completely characterized} by the existence of the generalized calibration $\Theta_P$ which also identifies the supersymmetric configurations of D-branes and networks. Differently from the $\mathbb{R}^{1,3}$ case we have used only two background Killing spinors (the time-independent ones) to construct the total calibration $\Theta$. This is related to the fact that we have only two independent supersymmetry equations (\ref{susyconds}), while the other two (\ref{intcond}) come as integrability conditions and it has a natural interpretation if one considers AdS$_4$ in Poincar\'e coordinates as a warped product of flat $\mathbb{R}^{1,2}$ times an internal direction $z$. Indeed, the two
independent time-independent supersymmetries correspond to the minimal supersymmetries on $\mathbb{R}^{1,2}$, while the other (time-dependent) two correspond to superconformal supersymmetries in the full $\caln=1$ superconformal algebra which is required to have a consistent supersymmetric AdS compactification.

Let us now allow for varying boundary
conditions and consider the case of a space-filling D-brane first. After we introduce
the boundary D-brane, the network of space-filling
D-brane and boundary domain wall is supersymmetric if and only if it calibrates \eqref{totcalp} where the relevant part is of course $\Theta_{P}^{\text{(sf)}}$.
The domain wall component will then calibrate the second term in \eqref{totcalpsf} and the $\d_H$ of this term will indeed produce the first term on the right-hand side of
\eqref{modpot} as promised. If we consider the energy difference of two such networks, this term, once integrated, will correspond to the energy of a domain wall at  $\partial \text{AdS}_4$ and stretching between the two space-filling D-brane configurations in the internal space. It can be considered as the energy cost associated to changing the boundary conditions. More complicated networks can be studied along the lines described in \cite{jarahluc}.

Analogously for domain walls the solution to the energy problem comes from the second term on the right-hand side in \eqref{totcalpdw}.
The structure of $\Theta_P$ allows moreover to identify the geometry of the possible domain walls in the AdS$_4$ part. Indeed, suppose that a D$p$-brane domain wall wraps a submanifold $\Pi$ in AdS$_4$ and a cycle $\Sigma$ in the internal space and let us focus on the part of $\Theta_P$ relevant for the domain wall component. The calibration condition can be split into a four-dimensional and an internal part
\begin{subequations}
\begin{align}
\sqrt{-g_{\text{AdS}}|_\Pi} \, \d^2 \sigma & = e^{i\alpha} (e^{\frac{3z}{R}}\d x^1\wedge\d x^2 - i e^{\frac{2z}{R}}\d x^1\wedge \d z) \, , \\
\sqrt{g|_\Sigma + \calf} \, \d^{p-2} \sigma & = \left[e^{i(\theta-\alpha)} \Psi_2 \wedge e^\calf\right]_{\text{top}} \, ,
\end{align}
\end{subequations}
for some phase $e^{i \alpha}$. For the four-dimensional part it follows
\eq{
\Im \left[e^{i\alpha} (e^{\frac{3z}{R}}\d x^1\wedge\d x^2 - i e^{\frac{2z}{R}}\d x^1\wedge \d z) \right] = 0 \, ,
}
from which we extract the profile
\eq{
\label{DWshape}
\begin{split}
\sin \alpha = 0 & : \quad z = c  \qquad \text{or} \\
\sin \alpha \neq 0 & : \quad \frac{\d x^2}{\d z} = \text{cotan} \, \alpha \; e^{-\frac{z}{R}} \Rightarrow x^2 = -R \, \text{cotan} \, \alpha \; e^{-\frac{z}{R}} + c \, ,
\end{split}}
with $c$ an integration constant. These profiles are basically the same as in \cite{kr}, but now in AdS$_4$ instead of AdS$_5$. The D-brane embeddings in AdS$_5 \times S^5$ of that paper were indeed shown to be supersymmetric in \cite{mtaylor}.  Through $e^{i\alpha}$ the profile
that a D-brane wraps in the four dimensions depends on the phase of the calibration form in the internal space.

A last remark  follows from the observation that AdS$_4$ in Poincar\'e coordinates can
be seen as a warped product of $\mathbb{R}^{1,2}$ with a radial coordinate. In this way one can consider compactifications to AdS$_4$ on a
six-dimensional internal space as compactifications to flat Minkowski $\mathbb{R}^{1,2}$ on a seven-dimensional internal space. For supersymmetric
compactifications the internal space has a $G_2 \times G_2$-structure which is built out of the $SU(3)\times SU(3)$-structure of the original
six-dimensional internal space. This leads to a realization of generalized Hitchin flow, which is discussed in appendix \ref{genhitchin}.

\subsection{Supersymmetry and energetics in global coordinates}

In the previous subsection we have described the total calibration for AdS$_4$ compactifications in Poincar\'e coordinates.
The reader might remark that it would be nice to also have a description of these total calibrations in global coordinates
since in contrast to Poincar\'e coordinates which only describe a patch of AdS$_4$, global coordinates parameterize the whole space.
Unfortunately the general procedure described in appendix \ref{susycalgen} cannot be applied straightforwardly
because the Killing spinors in global coordinates are time-dependent (see e.g.~\cite{bf} for explicit expressions) and thus do not fulfill a starting assumption of the analysis of appendix \ref{susycalgen}. It might seem surprising that we can describe
calibrations in one coordinate system and not in another. The basic reason is that so far our formalism of calibrations required the existence and choice of a global time coordinate and can only describe static configurations without electric world-volume gauge fields. An extension of the formalism to the non-static case would require a much more subtle treatment of the energy and we leave it for future work. In any case, it turns out that Poincar\'e and global coordinates are related by a time-dependent coordinate transformation so that the calibrated configurations, which are static in the one, are generically time-dependent in the other.

Our inability of applying the procedure of appendix \ref{susycalgen} to write a total calibration in global coordinates may suggest
that the arguments presented in subsection \ref{sketchsol}, explaining how to interpret the stability of supersymmetric D-branes,
are only valid in Poincar\'e coordinates.
Fortunately, we can still propose a candidate for a total calibration in global coordinates
even though it is not constructed from the supersymmetry Killing spinors. Let us take the (unwarped) AdS$_4$ metric in global coordinates
\bea
\d s^2_{(4)}=\frac{R^2}{\cos^2\!\rho}(-\d t^2 +\d\rho^2+\sin^2\rho\,\d \Omega_2)\ ,
\eea
where $0\leq\rho<\pi/2$ is the radial coordinate, $\d \Omega_2$ is the metric of a two-sphere  of radius one (parameterized by $(\theta,\phi)$)
and the boundary $\mathbb{R}\times S^2_{(\infty)}$ is located at $\rho=\pi/2$. The AdS space-like  volume form is thus given by
\bea
\text{vol}_3=\frac{R^4\sin^2\!\rho}{\cos^4\!\rho}\,\d \rho\wedge \text{vol}(S^2)\ ,
\eea
where $\text{vol}(S^2)=\sin\theta \,\d\theta\wedge\d \phi$.
In addition it is useful to introduce the following two- and one-dimensional volume elements
\bea
\text{vol}_2^{(1)}=\frac{R^3\sin\rho}{\cos^3\!\rho}\d\phi\wedge \d\rho\, ,\quad
\text{vol}_2^{(2)}=\frac{R^3\sin^2\!\rho}{\cos^3\!\rho}\,\text{vol}(S^2)\, , \quad
\text{vol}_1 = R^2\frac{\sin\rho}{\cos^2\!\rho}\d\phi \, ,
\eea
which correspond respectively to the volume form of the plane at $\theta=\pi/2$, of the two-sphere at constant radius $\rho$
and of their intersection. Then consider the polyform
\bea\label{totcalglob}
\Theta_G=\Theta^{\text{(sf)}}_G+\Theta^{\text{(DW)}}_G\, ,
\eea
with
\bea
\Theta^{\text{(sf)}}_G&=&\text{vol}_3\wedge \omega^{\text{(sf)}} +\chi_{2}\wedge \omega^{\text{(DW)}}_{\varphi=\theta} \, , \cr
\Theta^{\text{(DW)}}_G&=&\text{vol}_2^{(1)}\wedge \omega^{\text{(DW)}}_{\varphi=\theta-\frac{\pi}{2}}+\chi_{(1)}\wedge \omega^{\text{(string)}} \ ,
\eea
where we have introduced  $\chi_{(2)}\equiv R^3\tan^3\!\rho \; \text{vol}(S^2)$ and $\chi_{(1)}\equiv R^2\tan^2\!\rho\,\d\phi$ which are such that
\bea\label{diffG}
\d\chi_{(2)}=\frac{3}{R}\text{vol}_3\, ,\qquad \d\chi_{(1)}=-\frac{2}{R}\text{vol}_2^{(1)}\ ,
\eea
and for any two- respectively one-dimensional submanifold $\Pi$
\bea\label{ineq}
\chi_{(2)}|_{\Pi} \leq \big|\text{vol}_2^{(2)}|_{\Pi}\big| \leq \text{vol}(\Pi) \, , \qquad \chi_{(1)}|_{\Pi} \leq \big|\text{vol}_1|_\Pi \big| \leq \text{vol}(\Pi) \ .
\eea

We can see that $\Theta_G$ satisfies the two basic properties of an ordinary calibration.
Indeed, using the properties of the internal pure spinors one finds from
(\ref{diffG}) that $\d_H \Theta_G=-\text{vol}_3\wedge e^{4A}\tilde F$ and from (\ref{ineq})
that $\Theta_G$ satisfies the local algebraic bound
\bea\label{globound}
[\Theta_G|_\Sigma\wedge e^\calf]_{\rm top}\leq \cale_{\rm DBI}(\Sigma,\calf)\ .
\eea

Since we have not obtained  $\Theta_G$ from the background Killing spinors, we cannot immediately state that it
describes supersymmetric configurations. However, one can directly check that $\Theta_G$ still characterizes (homogeneous)
supersymmetric space-filling D-branes and straight radial domain walls ($\theta=\frac{\pi}{2}$), and can be safely used
to discuss their stability along the lines of subsections \ref{sketchsol} and \ref{poincarecord} because this analysis only
depends on the two basic properties we have just mentioned. These configurations correspond to the space-filling D-branes and only part
of the domain walls in Poincar\'e coordinates: only the ones
for which in \eqref{DWshape} $c=0$ and $\cos \alpha=0$ such that $x^2=0$. Indeed, after the transformation to global coordinates the spherical domain wall
of the first line of \eqref{DWshape} leads to a time-dependent configuration in global coordinates unless $z \rightarrow \infty$, and
the same applies for the interpolating domain walls ($\sin \alpha \neq 0, \cos \alpha \neq 0$). They are thus not found in the global coordinate
analysis. Indeed, the inequalities in (\ref{ineq}) can be saturated only at the spatial boundary $\rho=\pi/2$ so that
the calibration bound (\ref{globound}) can never be satisfied unless for straight radial domain walls or for spherical domain walls
at spatial infinity $\rho=\pi/2$ where the global time coincides with the Poincar\'e time.
$\Theta_G$ thus properly identifies the supersymmetry condition for the domain walls `at the end of the universe'.
In the same way also the strings `at the end of the universe' are described.

\section{Examples}
\label{examples}

\subsection{Background: $SU(3)$-structure vacua in IIA}
\label{examplesback}

As for the backgrounds, only type IIA ones with $SU(3)$-structure are studied in detail in the literature, so we will consider examples of calibrated
D-branes in this setting. An example of a type IIB vacuum with $SU(2)$-structure is given as the near-horizon geometry of a certain D-brane
configuration in \cite{tsimpisII}, while configurations with $SU(3)\times SU(3)$-structure presumably also exist, but no examples are known.

The $SU(3)$ vacua have been analysed in \cite{cvetic,tsimpis}. In this case we have $\eta^{(1)}_+=a\eta_+$ and $\eta^{(2)}_{+}=b\eta_+$ and we can take the two normalized pure spinors to be
\bea\label{SU(3)ps}
e^{i \theta} \Psi^+=e^{i\hat\theta}e^{iJ}\, ,\qquad \Psi^{-}=\hat\Omega=(a/b^*)\Omega \, ,
\eea
where $e^{i\hat\theta}=-ie^{i\theta}a/b$ and
\eq{
J_{mn} = i \eta_+^\dagger \hat{\gamma}_{mn} \eta_+ \, , \qquad \Omega_{mnp} = i \eta_-^\dagger \hat{\gamma}_{mnp} \eta_+ \,
}
define the $SU(3)$-structure. Plugging in (\ref{SU(3)ps}) into eqs.~(\ref{susyconds}) and (\ref{intcond}),
together with the (sourceless) Bianchi identity for $\hat{F}_0$
\eq{
\label{BiF0}
\d \hat{F}_0 = 0 \, ,
}
one gets that $\hat\theta,\Phi$ and $A$ must be constant, and we set $A=0$ without loss of generality\footnote{That the warp factor
should be always constant is rather mysterious especially if one should want to go beyond the probe limit and study the backreacted geometry
associated to, say, a localized supersymmetric D6-brane. One would expect this geometry to have a non-constant warp factor.}.
To further solve the supersymmetry conditions \eqref{susyconds} and \eqref{intcond}
it will be convenient to introduce the $SU(3)$ torsion classes $W_i$ as follows \cite{chiossal}
\begin{subequations}
\begin{align}
\d J & = -\frac{3}{2} \Im (\bar{W}_1 \hat\Omega) + W_4 \wedge J + W_3 \, \\
\d \hat\Omega & = W_1 \, J \wedge J + W_2 \wedge J + \bar{W}_5\,  \wedge \hat\Omega .
\end{align}
\end{subequations}
We find then
\bea\label{nk}
W_1 = -\frac{4}{3R}\cos\hat\theta\, , \qquad W_2 \,\, \text{real} \, , \qquad W_3 = W_4 = W_5=0 \, ,
\eea
which is called a {\em half-flat} geometry, and for the RR-fluxes
\bea\label{nksusy}
&&H=\frac{2\sin\hat\theta}{R}\Im\hat\Omega\, \quad m\equiv\hat F_{(0)}=\frac{5\sin\hat\theta}{R}e^{-\Phi}\, , \quad\hat F_{(2)}=-\frac{\cos\hat\theta}{3R}e^{-\Phi}J+e^{-\Phi} W_2\ ,\cr &&\hat F_{(4)}=\frac{3\sin\hat\theta}{2R}e^{-\Phi}J\wedge J\, ,\quad \hat F_{(6)}=\frac{\cos\hat\theta}{2R}e^{-\Phi}J\wedge J\wedge J\ .
\eea
It turns out that next to the equation of motion for $H$ also
the Bianchi identity for $\hat{F}_{(4)}$ automatically follows from the supersymmetry equations. The first was shown for general structure in \cite{paultsimpis}
for the Minkowski case and we extend the proof in appendix~\ref{susycondgen} to a general D-calibrated background with time-like Killing vector so
that in particular it also holds for AdS$_4$-compactifications.  The second also makes sense since
a supersymmetric D2-brane, which would source this Bianchi, is incompatible with $SU(3)$-structure.
The only non-trivial Bianchi identity is then the one for $F_{(2)}$, which we find from \eqref{beomRR} to be
\eq{
\d\hat F_{(2)}+mH= - j_3 \ ,
}
where $j_3$ is the source related to supersymmetric D6-branes and O6-planes. This leads to
\eq{
\label{F2B}
\frac{1}{3 R^2} \left( -2 \cos^2\hat\theta + 30 \sin^2 \hat\theta \right) \Im\hat\Omega + e^{-\Phi} \d W_2 = - j_3 \, .
}
Let us focus on the $(3,0)\oplus(0,3)$-part of this equation. From $W_2 \wedge \hat\Omega=0$ one can show that
\eq{
\d W_2 = \frac{1}{4} (W_2)^2 \, \Im \hat \Omega + (2,1) + (1,2) \, ,
}
with $(W_2)^2=\frac{1}{2}\,W_{2mn} W_2^{mn}$. Since the $(3,0)\oplus(0,3)$-part of the left-hand side of \eqref{F2B}
is non-singular, it follows that $j_3$ must correspond to all smeared sources. While it is puzzling that
we are not able to introduce localized sources, this is of course consistent with our earlier observation
that the warp factor has to be constant. Now, for supersymmetric and thus calibrated sources
we have
\eq{
j_3 = \sum_{l \in \text{sources}} c_{l} \, \Im \hat \Omega + (2,1) + (1,2) \, ,
}
where $c_l>0$ for a D6-brane and $c_l<0$ for an O6-plane. It follows that in the absence of O6-planes we have the bound
\eq{
|\sin\hat\theta| \le \frac{1}{4} \, .
}
AdS$_4$ compactifications can circumvent the no-go theorem of \cite{nogo} so that it should indeed be possible to find backgrounds that
satisfy all the equations of motion without introducing orientifold sources. As far as we know however examples without O6-planes
are only known for $W_2=0$ \cite{cvetic}, saturating the above bound (if there are also no smeared D6-branes) and leading to nearly K\"ahler geometry.
Since they also provide the easiest class of
solutions for the background we will present examples of calibrated D-branes in this setting.

Before we introduce the nearly K\"ahler manifolds in more detail, let us note
that on the other hand it is possible to introduce smeared O6-plane sources that are tuned such that they exactly compensate
the first term in \eqref{F2B}. One can then take $W_1=W_2=0$ and end up with a Calabi-Yau manifold. It was proposed in
\cite{acharya} that this is the underlying geometry for the model with all moduli stabilized of \cite{dewolfe}. However, there
does not seem to be any other compelling reason to tune the sources in this way except that it reduces to a Calabi-Yau analysis,
which is well understood.

Nearly K\"ahler manifolds are such that
their cone
\eq{
\d s^2_7 = \d u^2 + u^2 (w_1)^2(\d s^2_6) \, ,
}
is a $G_2$-holonomy manifold \cite{chiossal}. We have allowed for an extra constant warp factor $w_1$ to be determined in a moment.
Indeed, defining the associative and coassociative forms as follows
\begin{subequations}
\begin{align}
\label{assform}
\phi & = \d u \wedge u^2 (w_1)^2 J + u^3 (w_1)^3 \Im \hat\Omega \, , \\
\label{coassform}
\star_7 \phi & = \frac{1}{2} u^4 (w_1)^4 J \wedge J - \d u \wedge u^3 (w_1)^3 \Re \hat\Omega \, ,
\end{align}
\end{subequations}
one finds $\d_7 \phi = \d_7 (\star_7 \phi) = 0$ iff
\begin{subequations}
\begin{align}
& \d \Im \hat\Omega = 0 \, ,  & & \d J \wedge J =0 \, , & &\\
& \d  J = 3 w_1 \Im \hat\Omega \, ,
& & \d \Re \hat \Omega = - 2 w_1 J \wedge J \, . & &
\end{align}
\end{subequations}
Comparing with \eqref{nk} we find that we should take $w_1 = -(1/2) W_1 = (2/3R) \cos \hat\theta$. We remark that this is not
the same structure as the $G_2 \times G_2$-structure of \eqref{g2g2}. The latter, for which the RR-fields spoil the holonomy as in \eqref{int1}, is defined
in general, while only for $W_2=0$ we can define a $G_2$-{\em holonomy} (and then only if we define it differently as we would in general, namely
with the constant warping $w_1$). This could be compared with a Sasaki-Einstein geometry for supersymmetric compactifications to AdS$_5$, of which the cone has $SU(3)$-holonomy. This is also not the most general case \cite{beyondSE}.

The only homogeneous nearly K\"ahler manifolds in six dimensions
are \cite{nearlyK}
\begin{subequations}
\begin{align}
S^6 & \simeq \frac{G_2}{SU(3)} \, ,\\
S^3 \times S^3 & \simeq SU(2) \times SU(2) \, ,\\
\mathbb{CP}^3 & \simeq Sp(2)/SU(2)\times U(1) \, ,\\
F(1,2) & \simeq SU(3)/U(1)\times U(1) \, .
\end{align}
\end{subequations}
There are no non-homogeneous examples known.
In the following subsection we will each time present the D-brane example
for a general nearly K\"ahler background and then make it explicit for the special cases of $S^6$ and $S^3 \times S^3$, so let us present these geometries in some more detail.
$S^6$ is most easily described through its cone, which is
seven-dimensional flat space considered as the space
of imaginary octonions \cite{baez}. We can define the associative form as follows
\eq{
\phi (x,y,z) = G(x, y \cdot z) \, ,
}
where $G$ indicates the flat seven-dimensional metric, $x,y$ and $z$ are imaginary octonions and $\cdot$ is the octonionic product. In coordinates we can take
\begin{multline}
\phi = \d u^1 \wedge \d u^2 \wedge \d u^4 + \d u^1 \wedge \d u^5 \wedge \d u^6 + \d u^1 \wedge \d u^3 \wedge \d u^7 + \d u^2 \wedge \d u^3 \wedge \d u^5 \\
+ \d u^2 \wedge \d u^6 \wedge \d u^7 + \d u^3 \wedge \d u^4 \wedge \d u^6 + \d u^4 \wedge \d u^5 \wedge \d u^7 \, .
\end{multline}
The nearly K\"ahler space $S^6$ can then be taken to be the unit sphere. It can be shown that $J_z$, the right multiplication by the imaginary unit octonion $z$, induces a linear transformation on the tangent space $T_z S^6$, which moreover satisfies $J_z^2= - \bbone$. We can use it to define the complex structure and associated two-form
\eq{
(w_1)^2 J(x,y) = G(x,J_z y) = G(x,y \cdot z) = \phi(x,y,z) \, ,
}
where $x,y \in T_z S^6$. It follows that, if $z$ is given by the coordinates $(u^i)$,
\eq{
(w_1)^2 J = \iota_{u^i \partial_i} \phi \big|_{S^6} = \frac{1}{2} u^i \phi_{ijk} \d u^j \wedge \d u^k\big|_{S^6}\, ,
}
where as indicated we have to take the pullback to $S^6$. The $(3,0)$-form $\hat\Omega$ we find as follows
\bea\label{omegas6}
(w_1)^3 \hat\Omega = \left[ -\iota_{u^i \partial_i}(\star_7 \phi) + i \phi \right]\big|_{S^6} \, .
\eea

On $S^3 \times S^3$ there exists a unique left-invariant nearly K\"ahler structure \cite{nearlyK}, which is
given by
\begin{subequations}
\label{S3S3}
\begin{align}
\label{S3S3J}
J & = - \frac{1}{6 \sqrt{3} (w_1)^2} \left( e^1 \wedge f^1 + e^2 \wedge f^2 + e^3 \wedge f^3 \right) \, , \\
\label{S3S3O}\hat\Omega & = \frac{1}{(3 w_1)^3} \left\{ e^1 \wedge e^2 \wedge e^3 + f^1 \wedge f^2 \wedge f^3
- \frac{1}{2} \left( e^1 \wedge f^2 \wedge f^3 + f^1 \wedge e^2 \wedge f^3 + f^1 \wedge f^2 \wedge e^3\right) \right. \nonumber \\
& - \frac{1}{2} \left( f^1 \wedge e^2 \wedge e^3 + e^1 \wedge f^2 \wedge e^3 + e^1 \wedge e^2 \wedge f^3\right) \nonumber \\
& + i \sqrt{3} \left[
\frac{1}{2} \left( e^1 \wedge f^2 \wedge f^3 + f^1 \wedge e^2 \wedge f^3 + f^1 \wedge f^2 \wedge e^3\right) \right. \nonumber \\
& \left. \left. - \frac{1}{2} \left( e^1 \wedge e^2 \wedge f^3 + e^1 \wedge f^2 \wedge e^3 + f^1 \wedge e^2 \wedge e^3\right)
\right]
\right\} \, ,
\end{align}
\end{subequations}
where $e^i$ are the left-invariant Maurer-Cartan forms for the first $S^3$, $\d e^1=e^2 \wedge e^3$ and cyclic, and $f^i$
the ones for the second, $\d f^1 = f^2 \wedge f^3$ and cyclic.

\subsection{Examples of calibrated D-branes}

First we provide examples of calibrated D-branes on these nearly K\"ahler geometries that are space-filling from the four-dimensional point of view,
then we come to the domain walls. The reader can find an overview of all the calibrated D-branes we will describe in table \ref{caltable}.


\subsubsection*{Space-filling D-branes}

A space-filling D6-brane has to calibrate $\Re \hat{\Omega}$. More specifically the F-flatness condition $J|_\Sigma=\calf=0$ implies that the D-brane
wraps a Lagrangian submanifold in the internal space. {}From \eqref{fimpliesd}, which says that the F-flatness implies the D-flatness, follows that this manifold
is automatically special Lagrangian i.e.\ $\Im \hat{\Omega}|_\Sigma = 0$. In this context of Lagrangian submanifolds of nearly K\"ahler manifolds this was already noted in \cite{ivanpap}.

A concrete example on $S^6$ is the equatorial $S^3 \subset S^6$ with $u^1=u^2=u^4=0$. One can easily verify the special Lagrangian conditions directly or alternatively
note from \eqref{coassform} that the intersection of $S^6$ with a radially extended coassociative cycle in $\mathbb{R}^7$ leads to a special Lagrangian cycle.
As shown in \eqref{soleom} a D-brane wrapping a submanifold like this solves its equations of motion, meaning its energy is stationary under small variations, but (naively) it is not at minimal energy. This implies that there are tachyonic modes.
And moreover, an $S^3$ inside $S^6$ is homologically trivial so one might expect that it would just shrink to zero. In section \ref{adstot} however we showed
the stability in general. Let us present an explicit check for our example. Suppose we consider a perturbation $\delta u^1$. {}From the Dirac-Born-Infeld action we find that the action for $\delta u^1$ up to quadratic order is proportional to
\eq{
- \frac{1}{2} g^{\mu\nu} \partial_{\mu} \, \delta u^1 \partial_{\nu} \, \delta u^1 + \frac{3}{2} (w_1)^2 (\delta u^1)^2 \, ,
}
from which we find
\eq{
m^2 = - 3 (w_1)^2 = - \frac{4}{3R^2} \cos^2 \! \hat{\theta}  \, ,
}
satisfying the Breitenlohner-Freedman bound $m^2 > - \frac{9}{4R^2}$ \cite{bf} as predicted by the general discussion in \ref{sketchsol}.
So they do not signal an instability.

On $S^3 \times S^3$ we see immediately from \eqref{S3S3} that a D6-brane wrapping one of the $S^3$ factors --- homologically non-trivial this time --- is calibrated.

A calibrated D8-brane wraps a codimension-one submanifold in the internal space.  The real problem is finding the world-volume gauge field so that $\d \calf=H|_\Sigma$
and the F-flatness condition
\eq{
(i J|_\Sigma + \calf)^2 = 0\ ,
}
which is equivalent to the condition for having a coisotropic D-brane \cite{coisotropic},
is satisfied. Again the D-flatness $\Im \hat\Omega|_\Sigma \wedge \calf = 0$ then follows automatically. Since $H=\tan \hat\theta \, \d J$, we have a globally defined $B$-field $B=\tan \hat\theta J$, and we can put $\calf=\tan\hat\theta J|_\Sigma+{\text F}$, with ${\text F}$ a necessarily closed $U(1)$ world-volume field-strength.
It is impossible to have supersymmetric space-filling D8-branes with ${\text{F}}=0$ since the
F-flatness condition would then reduce to
 \bea
 (J\wedge J)|_\Sigma=0\ ,
 \eea
 which cannot be satisfied. On the other hand, with a non-trivial world-volume field-strength ${\rm F}$ we have the conditions
 \bea
 (J \wedge J)|_\Sigma = \cos^2 \! \hat\theta \, \text{F}\wedge \text{F}= -{\rm cotan} \, \hat\theta\,J|_\Sigma \wedge \text{F} \ ,
 \eea
which would leave room for non-trivial configurations. Unfortunately, we did not find any solutions nor were we able to show that a solution is impossible.\footnote{Examples of coisotropic D-branes are very few. As far as we know as for the five-dimensional codimension-one coisotropic D-branes explicit solutions have only been found
on a torus with $H=0$ \cite{marchesanoco}.}


\begin{table}
\begin{center}
\begin{tabular}{|cc|ccc|}
\hline
D-brane & Type & Internal cycle & $\calf=\tan\hat\theta J|_\Sigma+\text{F}$ & $e^{i\alpha}$ \\ \hline
D6 & sf & \rule[1.1em]{0pt}{0pt} SLAG & $\text{F}=0$ & NA \\
D8 & sf & codimension-1 & no solution found & NA \\ \hline
D2 & DW & point & $\calf=0$ & \rule[1.2em]{0pt}{0pt} $e^{i\alpha}=e^{i \hat\theta}$ \\
D4 & DW & complex 2-cycle & $\calf= - \text{cotan}(\hat\theta - \alpha) \, J|_\Sigma$ & fixed by $\int_\Sigma \text{F}\in\mathbb{Z}$\\
D6 & DW & not possible & / & / \\
D8 & DW & M & F $(1,1)$ and primitive & $e^{i\alpha}(\hat\theta)$ \\ \hline
\end{tabular}
\caption{Overview of calibrated D-branes in nearly K\"ahler type IIA $SU(3)$-structure vacua.\label{caltable}}
\end{center}
\end{table}

\subsubsection*{Domain wall D-branes}

Also for domain walls, it is useful to rewrite the calibration condition (\ref{cc1}) as a pair of conditions (\ref{altersusy}), which we
will still refer to as F-flatness and D-flatness conditions in analogy to the case of space-filling branes. However, now F-flatness does {\em not} automatically imply D-flatness so that the two conditions have to be checked separately.

For all kinds of domain walls the F-flatness conditions read
\bea\label{Fdw}
\big[\hat\Omega|_\Sigma\wedge e^\calf\big]_{\text{top-1}}=0\, , \qquad \big[\iota_X\hat\Omega|_\Sigma\wedge e^\calf\big]_{\text{top}}=0\quad \forall X\in T_M\ ,
\eea
which can be rephrased as the requirement that they
should wrap an ``almost'' complex cycle in the internal space, i.e.\ a cycle for which the tangent space
in every point is stable under the almost complex structure, and $\calf$ is of type (1,1).  As the complex structure is not integrable it is not evident that such cycles exist, as the tangent spaces might not be integrable. So let us discuss the different cases separately. A D2-brane obviously always satisfies the F-flatness conditions (\ref{Fdw}) since it is point-like in the internal space.

On the other hand, calibrated D6-brane domain walls do not exist since
there is no almost complex 4-cycle. We can show this from the following property of the Courant bracket \cite{gualtieri}
\eq{
[\mathbb{X},\mathbb{Y}] \cdot \Psi = \mathbb{X} \cdot \mathbb{Y} \cdot \d_H \Psi  \, ,
}
with $\Psi$ a pure spinor and $\mathbb{X},\mathbb{Y} \in \Gamma(L)$. Applied to the case at hand, it follows  for $v,w \in \Gamma(T_M^{0,1})$
that
\eq{
[v,w]|_{\Gamma(T_M^{1,0})}^i = w_1 \, \hat{\bar{\Omega}}{}^i{}_{jk} v^j w^k \, ,
}
so that for an integrable tangent space two complex coordinates will always induce the third.

For D4-branes there are generically solutions to (\ref{Fdw}). On $S^6$ a concrete example of a D4-brane satisfying (\ref{Fdw}) would be the equatorial $S^2 \subset S^6$ with $u^3=u^5=u^6=u^7=0$. Indeed the F-flatness condition reduces then to $(\iota_X\hat\Omega)|_{S^2}=0$ for any $X\in T_{S^6}$ and is satisfied since
plugging in \eqref{omegas6} we find
\bea
(\iota_X\hat\Omega)|_{S^2}=i(w_1)^{-3}[\iota_X\phi|_{S^6}]|_{S^2}=i(w_1)^{-3}[\iota_X(\d u^1\wedge\d u^2\wedge\d u^4)|_{S^6}]|_{S^2}=0\ ,
\eea
which can be checked by introducing appropriate angular coordinates for $S^6$ and $S^2\subset S^6$. Alternatively, one can note that from \eqref{assform}
follows that the intersection of $S^6$ with a radially extended associative cycle in $\mathbb{R}^7$ leads to an almost complex cycle.
On $S^3 \times S^3$ an example of an almost complex two-cycle would be $S^1 \times S^1$ with the first $S^1$ equatorial in the first $S^3$ and the second $S^1$ the corresponding equator (by \eqref{S3S3J}) in the second $S^3$. Again, the condition $(\iota_X\hat\Omega)|_{S^1\times S^1}=0$ can be easily checked from the explicit form of $\hat\Omega$ given in (\ref{S3S3O}), taking into account that the pull-backs of $e^2, e^3, f^2$ and $f^3$ to $S^1\times S^1$ vanish.

Regarding D8-branes, they are always almost-complex since they completely fill the internal space, and thus (\ref{Fdw}) reduces to the condition that $\calf$ must be $(1,1)$. In general nearly K\"ahler backgrounds we can always solve this condition together with the modified Bianchi identity $\d\calf=H|_\Sigma$ by putting $\calf=\tan\hat\theta J|_\Sigma+\text{F}$, where $\text{F}$ is a necessarily closed $(1,1)$  $U(1)$ world-volume field-strength.

Rests us to analyse the D-flatness condition, which reads
\eq{
\label{DDW}
\Im \big[ e^{i(\hat{\theta}-\alpha)}e^{i J|_\Sigma + \calf}\big]_{\text{top}} = 0 \, ,
}
for a constant phase $e^{i \alpha}$ that also determines the shape of the domain wall in the four dimensions as in \eqref{DWshape}. For a D2-brane we find $e^{i\alpha}=e^{i \hat\theta}$. For the other cases, we can again use the splitting  $\calf=\tan\hat\theta J|_\Sigma+\text{F}$, with $(1,1)$ by F-flatness, and conclude that D-flatness requires
\begin{subequations}
\begin{align}
D4 & :\qquad \text{F}= - [\tan\hat\theta+\text{cotan}(\hat\theta - \alpha)] \, J|_\Sigma \, , \quad {e^{i \alpha}} \, \text{classically free} \, , \label{d4d}\\
D8 & :\qquad \text{F\ primitive} \, ,\qquad{e^{i \alpha}} \, \text{fixed} \, .
\end{align}
\end{subequations}
For the D8 the relation between $\alpha$ and $\hat\theta$ depends on $\text{F}$. For example, if we take ${\rm F}=0$ then $\cos (2 \hat\theta - \alpha) =0$. On the other hand, for D4-branes $\alpha$ is classically unfixed, but at the quantum level must satisfy a constraint coming from the quantization of the $U(1)$ field-strength $\text{F}$, which imposes that
\bea
\int_\Sigma \text{F}=n\in\mathbb{Z}\ .
\eea

\section{Conclusions}

In this paper we have studied several aspects of supersymmetry and stability of D-branes on flux compactifications of type II theories to AdS$_4$ space-time. Most of the arguments rely on the existence of appropriate background generalized calibrations in the sense of \cite{gencal,luca1,jarahluc}. They identify supersymmetric D-branes and must obey differential conditions which turn out to be equivalent to the requirement that the background is supersymmetric. This deep relation between D-brane and background structures allows us to give general arguments proving the classical stability (even under large deformations) of supersymmetric D-branes, some of them wrapping trivial cycles in the internal six-dimensional space. Our results can be seen as a geometrical ten-dimensional realization in string theory of the results of \cite{bf,freedman} for supersymmetric field theories in AdS$_4$. It would be nice to extend our results to include the coupling to the closed string sector described in e.g.~\cite{palti,granaN2,grimm,effective,cassbil,kp}.

Even though we have used the classical theory, it should be possible to incorporate quantum perturbative and non-perturbative corrections along the lines of \cite{effective}  without spoiling the main points  of our discussion. Furthermore, we have focused on the case of four-dimensional compactifications having phenomenology as main motivation, but many arguments are more general. Indeed, in appendix \ref{susycalgen} we have proved the relation between the conditions for general static supersymmetric backgrounds (admitting at least a static Killing spinor),  of which AdS$_4$-compactifications are just a special case, and the differential conditions for calibrations associated to static supersymmetric D-branes.

However, there are many interesting D-brane configurations that are supersymmetric (possibly with respect to a  non-static Killing spinor) but not static because they either are time-dependent, like giant gravitons \cite{giantgraviton}, or they have electric world-volume gauge fields, like the F1-D3 intersection (the electric BIon) \cite{bion}. It would be interesting to extend the theory of calibrations to these configurations \cite{hjsmith}.

The extension to more general backgrounds is also interesting for applications to the AdS/CFT correspondence. For example the addition of branes filling the AdS part of the ten-dimensional geometry corresponds to the addition of flavours to the dual conformal field theory (see e.g.~\cite{KK}). This generically destroys the conformal invariance in the resulting gauge theory and indeed, after taking into account the backreaction of the flavour branes in the dual geometry,  the AdS space should be substituted by a geometry encoding the non-trivial renormalization group flow of the dual gauge theory (see e.g.~\cite{malda,hashimoto} for a discussion considering localized sources and \cite{benini} for more recent examples using smeared D-branes). This effect seems puzzling in the context of phenomenologically relevant flux compactifications to AdS$_4$ where the introduction of D-branes and orientifolds preserving the background supersymmetry is not expected to modify the AdS$_4
 $ vacuum geometry, like it is indeed the case in compactifications to flat space.

We see this puzzle about the backreaction directly in the examples of IIA $SU(3)$-structure backgrounds, discussed in section \ref{examplesback},  where the supersymmetry conditions (together with the Bianchi identity for $\hat{F}_0$) force the warp factor to be constant. This means that, while {\em probe} supersymmetric space-filling D-branes are possible, and indeed we constructed explicit examples in the special case of nearly K\"ahler backgrounds, a problem arises when one tries to consider the backreaction of such a D-brane if it is fully localized. Indeed, close to the D-brane the backreacted geometry should be similar to the backreaction of the D-brane in flat space, so that we expect a non-constant warp factor. It is possible that the structure of the backreacted geometry gets deformed to a genuine $SU(3)\times SU(3)$-structure, which presumably does allow non-constant warp factor. This would however also mean that the generators of the preserved internal supersymmetry change, which is puzzling for the backreaction of D-branes that were already supersymmetric with the background as probes. Another possibility is that taking into account stringy corrections might allow for non-constant warp factor. We hope to report on this problem in a future publication.

\vspace{0.9cm}

\section*{Acknowledgements}
We thank G.~Barnich, F.~Bigazzi, G.~Bonelli, J.~Gauntlett, M.~Petrini, P.~Smyth, D.~Tsimpis, A.~Zaffaroni and M.~Zagermann for useful discussions. The work of P.~K.\ is supported by the German Research Foundation (DFG) within the Emmy-Noether-Program (Grant number ZA 279/1-2). L.~M.\ is presently supported by the DFG cluster of excellence `Origin and Structure of the Universe' and during the bulk part of this work he was at the ITF of the K.~U.\ Leuven and supported in part by the Federal Office for Scientific, Technical and Cultural Affairs through the ``Interuniversity Attraction Poles Programme -- Belgian Science Policy" P5/27 (2006) and P6/11-P (2007), and by the European Community's Human Potential Programme under contract MRTN-CT-2004-005104 `Constituents, fundamental forces and symmetries of the universe'. L.~M.\ wishes to acknowledge also the Galileo Galilei Institute for Theoretical Physics for hospitality and the INFN for partial support.

\vspace{0.9cm}





\begin{appendix}

\section{Supersymmetry and calibrations: a general discussion}
\label{susycalgen}

We would like to discuss the relation between the supersymmetry of a certain supergravity vacuum and the fact that it is automatically equipped with a calibration that allows to characterize its supersymmetric D-branes. In the absence of fluxes, this relation is well known \cite{calold} and relies on the physical property that supersymmetric branes are naturally volume-minimizing. The way to include (some of the) background fluxes was indicated in \cite{papa}, where a calibration was required to minimize the brane energy rather than the volume. See for example \cite{gauntlettcalbulk} for work applying/developing this idea.  This work also indicates that the relation
is not restricted to D-branes, see also \cite{paultsimpis} for a calibration for space-filling NS5-branes.
However the prescription given in \cite{papa} does not allow to take into account world-volume fluxes. This limitation turns out to be problematic when considering D-branes on type II backgrounds with general fluxes, since the world-volume flux $\calf$ is related to the background $H$-field by the modified Bianchi identity $\d \calf=H|_\Sigma$.  The way to solve this problem was presented in \cite{gencal,luca1} (see also \cite{jarahluc}).

Before recalling our definition of  a generalized  calibration we have to discuss the structure of the type II backgrounds we are considering. First of all, we require that the background admits a globally defined time-like Killing vector so that it is possible to introduce an associated time $t$. Secondly, as we will recall later, calibrations allow to characterize the lower bound of the energy of a certain D-brane configuration in terms of its `topological' properties \cite{jarahluc}. Thus we consider a setting in which possible dynamically conserved charges, which are not  topological in nature and may enter the D-brane energy, are set to zero. A natural way to achieve this  is by requiring  both the space-time and the D-branes to be static.

We thus assume that the ten-dimensional space-time is topologically $X=\mathbb{R}\times \calm$, with coordinates $X^M=(t,y^m)$, and the ten-dimensional metric splits as
\bea
G\equiv G_{MN}\d X^M\d X^N=-e^{2A(y)}\d t^2 + g_{mn}(y)\d y^m\d y^n,
\eea
where $g$ is the metric on $\calm$.
The warp factor $A$ and the dilaton $\Phi$ depend only on the internal coordinates $y^m$, just as the $H$ field which we also assume to only have internal legs. Furthermore, grouping the RR field-strengths in $\d_H$-closed polyforms $F=\sum_{k}F_{(k)}$, where $k$ is even in IIA and odd in IIB,  they can be decomposed as
\bea
F=\d t\wedge e^A\tilde F +\hat F\ ,
\eea
where $\tilde F$ and $\hat F$ are polyforms on $\calm$ that do not depend on the time $t$. As in \cite{luca1} we use the conventions of \cite{lucadirac}, in which we have the following duality relation between the electric and magnetic RR fields
\bea
\tilde F=\sigma(\star_9 \hat F)\ ,
\eea
where $\sigma$ reverses the order of the indices of the form it is acting on. In this setting we will study static D-brane configurations with world-volume
gauge field $\calf$ purely along $\calm$ i.e.\ we do not consider electric world-volume gauge fields.

\subsection{Generalized calibrations: definition and main properties}
\label{reviewcal}

According to \cite{gencal,luca1} a {\em generalized calibration}\footnote{In the following we will omit the adjective `generalized', implicitly always assuming
the calibrations with both world-volume and background fluxes.} is given by a $\d_H$-closed polyform $\hat{\omega}$ on $\calm$ such that, for any static D-brane wrapping the generalized submanifold $(\Sigma,\calf)$  of $\calm$, one has
\bea\label{alg}
[\hat{\omega}|_\Sigma\wedge e^\calf]_{\rm top}\leq \cale(\Sigma,\calf)\ ,
\eea
where $\cale(\Sigma,\calf)$ is the energy density of the D-brane configuration. Choosing the electric component of the RR field-strengths $\tilde F$ as the fundamental one, one can decompose the RR gauge potentials $C=\sum_k C_{(k)}$ as $C=\d t\wedge e^A\tilde C$. Then, the D$p$-brane energy density $\cale(\Sigma,\calf)$ appearing in (\ref{alg}), which can be extracted from the D-brane action consisting of a Dirac-Born-Infeld and a Chern-Simons part, is given by
\bea
\cale(\Sigma,\calf)=e^{A-\Phi}\sqrt{\det(g|_\Sigma+\calf)}\,\d^p\sigma-e^A[\tilde C|_\Sigma\wedge e^\calf]_{\rm top}\ .
\eea
Note that the definition of the calibration $\hat{\omega}$ depends on the gauge choice for the RR potential $\tilde C$.  An alternative {\em gauge-invariant} definition is however possible, according to which the calibration is given  by a polyform $\omega$  such that
\bea
\label{alg2}
[\omega|_\Sigma\wedge e^\calf]_{\rm top}\leq e^{A-\Phi}\sqrt{\det(g|_\Sigma+\calf)}\,\d^p\sigma\ ,
\eea
and
\bea\label{integr}
\d_H\omega  =-e^A\tilde F\ .
\eea
Since $\tilde F=-e^{-A}\d_H(e^A \tilde C)$, the two definitions are obviously related by
\bea
\label{calform}
\hat{\omega}=\omega-e^A\tilde C\ .
\eea

We say that the space $\calm$ is {\em D-calibrated} if it is equipped with a calibration $\hat{\omega}$ (or $\omega$) as defined above. If this is the case, a generalized submanifold $(\Sigma,\calf)$ of $\calm$ is said to be {\em calibrated} with respect to $\hat{\omega}$ (or $\omega$) if at any point the inequality (\ref{alg}) (or equivalently (\ref{alg2})) is saturated, i.e.\ in terms of $\omega$:
\bea\label{calcond}
[\omega|_\Sigma\wedge e^\calf]_{\rm top}= e^{A-\Phi}\sqrt{\det(g|_\Sigma+\calf)}\,\d^p\sigma, \qquad \text{at any point}\in \Sigma\ .
\eea

As discussed in \cite{jarahluc}, one can define a proper boundary operator $\hat\partial$ acting on chains of generalized submanifolds of possibly different dimensions. Thus the theory naturally allows to treat networks of D-brane of different dimensions and gauge invariance requires a D-brane network to wrap a generalized cycle defined by $\hat\partial$. Furthermore, since the calibration $\hat{\omega}$ is $\d_H$-closed, it is easy to see that a calibrated D-brane network minimizes its energy inside its generalized homology class, and this `topological' minimal energy  is obtained by integrating the calibration $\hat{\omega}$ over any representative generalized cycle inside the corresponding generalized homology class. We refer to \cite{jarahluc} for more details and display here for simplicity the argument for a single D-brane.
Take $(\Sigma,\calf)$ a calibrated D-brane and $(\Sigma',\calf')$ any other D-brane in the same generalized homology class which means
there is a $(\Gamma,\tilde{\calf})$ such that $\partial \Gamma=\Sigma' - \Sigma$ and $\tilde{\calf}|_\Sigma=\calf$, $\tilde{\calf}|_{\Sigma'}=\calf'$. We find
\eq{
\label{calboundshow}
\cale(\Sigma',\calf') \ge \int_{\Sigma'} \hat{\omega}|_{\Sigma'} \wedge e^{\calf'} = \int_{\Sigma} \hat{\omega}|_{\Sigma} \wedge e^\calf + \int_\Gamma \d_H \hat{\omega}|_\Gamma \wedge e^{\tilde{\calf}}= \cale(\Sigma,\calf) \, ,
}
where we used \eqref{alg}, $\d_H \hat{\omega}=0$ and \eqref{calcond}. This demonstrates clearly that the $\d_H$-closedness is an essential part
of the definition of the calibration form and explains why the would-be calibrations of section \ref{susyd} have problems with stability.

To a generalized submanifold $(\Sigma,\calf)$ a generalized current $j_{(\Sigma,\calf)}$ in $\calm$ can be associated \cite{deforms}, which is defined
such that
\eq{
\int_\Sigma \phi|_\Sigma \wedge e^\calf= \int_{\calm} \langle \phi, j_{(\Sigma,\calf)} \rangle \, ,
}
for any polyform $\phi$. The cohomology of these currents is dual to the above introduced generalized homology.
The energy density $\cale(\Sigma,\calf)$ of the calibrated D-brane is then given by
\eq{
\cale(\Sigma,\calf)=\int_\calm e^{A-\Phi} \langle \hat{\omega} , j_{(\Sigma,\calf)} \rangle \, ,
}
and the calibration conditions can be expressed in terms of $j_{(\Sigma,\calf)}$ (see \cite{deforms} in the case of compactifications to
four-dimensional flat Minkowski space). One can analogously define a current $j^X_{(\Sigma,\calf)}$ for the same generalized submanifold, but now seen as a submanifold
of the total ten-dimensional space $X$. It can be decomposed as
\eq{
j^X_{(\Sigma,\calf)} = \d t \wedge e^A \tilde{\jmath}_{(\Sigma,\calf)} + j_{(\Sigma,\calf)} \, .
}
The restriction throughout the paper to static D-branes without electric world-volume gauge fields translates into the statement that
\eq{
\label{extsource}
\tilde{\jmath}_{(\Sigma,\calf)} = 0 \, .
}

Beyond the probe approximation, D-branes and orientifolds\footnote{In supergravity orientifolds are described in the same way as D-branes, except that their
tension is negative and $\calf$ vanishes.} act as sources for the RR-fields and change their equations of motion and Bianchi identities as follows (see e.g.~\cite{paultsimpis})
\eq{
\label{beomRR}
\d_H \hat{F} = - j_{\text{tot}} \, , \qquad \d_H(e^A \tilde{F}) = \tilde{\jmath}_{\text{tot}} = 0 \, ,
}
with $j_{\text{tot}} = \sum_{(\Sigma,\calf)_{\text{D}p}} T_{\text{D}p} \, j_{(\Sigma,\calf)_{\text{D}p}} - \sum_{(\Sigma)_{\text{O}p}} T_{\text{O}p} \,  j_{(\Sigma)_{\text{O}p}}$.
Furthermore it was shown in \cite{paultsimpis} that exactly if these sources are calibrated, supersymmetry of the background together with
the above source-corrected Bianchi identities and equations of motion of the form-fields implies the source-corrected Einstein and dilaton equations.

\subsection{Supersymmetry and generalized geometry of $\calm$}
\label{susycondgen}

Let us now discuss the relation with supersymmetry. As we will discuss, under mild conditions supersymmetry equips vacua with calibrations. Vacua that
satisfy these conditions will be called D-calibrated.

The supersymmetry is generated by two Majorana-Weyl Killing spinors $\epsilon_1$ and $\epsilon_2$ of opposite/same chirality in IIA/IIB.
One can construct two one-forms
\bea\label{1forms1}
V^{(1)}_M=\bar\epsilon_1\Gamma_M\epsilon_1\, , \qquad V^{(2)}_M=\bar\epsilon_2\Gamma_M\epsilon_2\ ,
\eea
and combine them into the one-forms $V^{(\pm)}=V^{(1)}\pm V^{(2)}$. Indicating with $V_{(\pm)}$ the corresponding vectors obtained by raising the index with the ten-dimensional metric $G$, one finds from the supersymmetry Killing spinor conditions (see also e.g.~\cite{hjsmith,superalgebra})
\bea\label{killing}
\call_{V_{+}}G=0\, , \qquad \d V^{(-)}=-\iota_{V_{(+)}}H\ .
\eea
These equations imply that $V_{(+)}$ is a Killing vector of the ten-dimensional metric and $H$-field (the second equation is actually stronger).

Eqs.~(\ref{killing}) are valid in full generality, without any restriction on the bosonic configuration.  However in order to proceed, for our purposes we restrict to the static 1+9 splitting described at the beginning of this appendix, even though the following steps may be repeated in a completely generic setting.  The gamma-matrices decompose as
\bea
\Gamma_{\underline{0}}=i\sigma_2\otimes\bbone\, ,\qquad \Gamma_{\underline{m}} = \sigma_1\otimes\gamma_{\underline{m}} \, , \qquad \Gamma_{(10)} = \sigma_3\otimes\bbone \, ,
\eea
where underlining indicates flat indices, $\gamma_{\underline{m}}$ are the nine-dimensional gamma-matrices, which we take real and symmetric, and ${\sigma}_i$ the standard Pauli matrices.
The supersymmetry generators split as
\bea
\epsilon_1 =  \left(\begin{array}{c} 1 \\ 0 \end{array}\right)\otimes \chi_1 \, , \qquad
\epsilon_2 = \left\{ \begin{array}{l}\left(\begin{array}{c} 0 \\ 1 \end{array}\right)\otimes\chi_2  \quad \text{(IIA)} \\ \\
\left(\begin{array}{c} 1 \\ 0 \end{array}\right)\otimes \chi_2 \quad \text{(IIB)}\end{array}\right.\ ,
\eea
where $\chi_{1,2}$ are real spinors on $\calm$. We indicate their norms with
\bea
\chi_1^T\chi_1=a^2\, , \qquad \chi_2^T\chi_2=b^2\ ,
\eea
and construct two one-forms on $\calm$ as follows
\bea\label{1forms2}
v^{(1)}_m=\chi^T_1\gamma_m\chi_1\, , \qquad v^{(2)}_m=\mp\chi^T_2\gamma_m\chi_2\quad \text{in IIA/IIB}\ .
\eea
The one-forms (\ref{1forms1}) on $X=\mathbb{R}\times \calm$ then look in 1+9 notation like
\bea
V^{(1)}=(e^Aa^2, -v^{(1)})\, , \qquad V^{(2)}=(e^Ab^2,-v^{(2)})\ .
\eea
Defining $v^{(\pm)}=v^{(1)}\pm v^{(2)}$ we find from (\ref{killing}) applied to our static configuration two sets of conditions. We first have the following geometrical properties for $\calm$
\bea\label{9eq1}
v_{(+)}(A)=0\, , \qquad \call_{v_{(+)}}g=0\, , \qquad \d v^{(-)}=-\iota_{v_{(+)}}H\ ,
\eea
while the second set relates the norms of the internal spinors $\chi_1$ and $\chi_2$ to the warp factor
\bea\label{9eq2}
\d[e^{-A}(a^2+b^2)]=0\, , \qquad \d[e^A(a^2-b^2)]=0\ .
\eea

The two internal spinors $\chi_1$ and $\chi_2$ can now be used to construct a real polyform of definite parity $\Psi$ associated by the Clifford map to the tensor product
\bea
\slashchar{\Psi}=\chi_1\otimes \chi_2^T\ .
\eea
Note that in nine dimensions a complete base for the real $16\times 16$ real matrices is given by all the $\gamma_{m_1\ldots m_p}$  with $p$ either even or odd. This means that $\slashchar{\Psi}$ corresponds to either an even or an odd polyform\footnote{The two possibilities are related by Hodge duality, as can be obtained from the `self-duality' $\slashchar{\lambda}=-\slashchar{\sigma(\star_9\lambda)}$ which is valid for any form $\lambda$ on $\calm$.}. As will become clear soon, we need to choose the even representative in IIA and the odd representative in IIB\footnote{This choice can for example be understood by requiring a natural ten-dimensional origin for $\Psi$, which can be defined as $\Psi_{m_1\ldots m_p}\sim \bar\epsilon_2\Gamma_{m_1\ldots m_p}\epsilon_1$.}:
\bea
\Psi=\left\{ \begin{array}{l}
\Psi_{\rm (even)}\quad\text{in IIA}\\
\Psi_{\rm (odd)}\quad\ \text{in IIB}\\
\end{array}\right.\quad .
\eea

{\sloppy A long, but straightforward calculation, which is similar to the analogous six-dimensional one of \cite{gmpt}, shows that the Killing spinor equations can be manipulated
into a condition for $\Psi$ in both IIA and IIB, which reads
\bea\label{int1}
\d_H(e^{-\Phi}\Psi)=\frac{1}{32}(a^2+b^2)\tilde F-\frac{1}{32}v^{(-)}\wedge \hat F-\frac{1}{32}\iota_{v_{(+)}}\hat F\ .
\eea}
\fussy
The integrability of this equation together with (\ref{9eq1}), (\ref{9eq2}) and \eqref{beomRR} implies that
\eq{
\call_{v_{(+)}} \hat F + \mathbb{V} \cdot j_{\text{tot}} = 0 \, ,
}
where we have defined the generalized vector $\mathbb{V}=(v_{(+)},v^{(-)})$. If all sources are calibrated, which implies \eqref{calsource} as we will see in
the next subsection, we find that the second term above is zero so that also $\call_{v_{(+)}}\hat F=0$. Using the Killing spinor equations it is furthermore possible to see that $v_{(+)}(\Phi)=0$. Thus $v_{(+)}$ generates a symmetry of the full background configuration.
One can immediately recognize in (\ref{int1}) a strong similarity with eq.~(\ref{integr}), thus suggesting a natural calibration form for D-branes
of the type described in subsection \ref{reviewcal}. To make this point more precise, we have to discuss the supersymmetry of probe D-branes in our background. This will allow us to clarify when our supersymmetric vacua turn out to be really D-calibrated.

But let us first note for completeness that the following relation also follows in a similar way from the background supersymmetry
\begin{multline}
-s \, \d \left[16 \, e^{-2 \Phi} \left(\chi_1 \otimes \chi_1^T - \chi_2 \otimes \chi_2^T \right)\right]_5 \\
= (a^2+b^2)(e^{-2 \Phi} \star_9 H) \mp 32 \, \left[\sigma(\hat{F}) \wedge e^{-\Phi} \Psi\right]_6 \, ,
\end{multline}
for IIA/IIB, and where $s$ is a sign defined by $\gamma_{\underline{1 \ldots 9}}=s \bbone$. The 5-form on the left-hand side
can presumably be used to construct the calibration form for an NS5-brane, although we will not study this
in detail in this paper since the action and supersymmetry of such a brane in the presence of world-volume
gauge fields is quite complicated. From the integrability of the above equation follows in a way similar to
\cite{paultsimpis} that if \eqref{cons} is satisfied the background supersymmetry implies the equation of motion for $H$.

\subsection{Introducing D-branes: when is $\calm$ D-calibrated?}

Let us now introduce a static probe D-brane in our supersymmetric background, filling the time direction and wrapping a generalized cycle $(\Sigma,\calf)$ in $\calm$. The usual $\kappa$-symmetry argument adapted to our case implies that a D$p$-brane is supersymmetric if and only if
\bea\label{kappa}
\gamma_{\text{D}p}(\calf)\chi_2=\chi_1\ ,
\eea
with
\bea
\gamma_{\text{D}p}(\calf)=\frac{1}{\sqrt{\det
(g|_\Sigma+{\cal
F})}}\sum_{2l+s=p}\frac{\epsilon^{\alpha_1\ldots\alpha_{2l}\beta_{1}\ldots\beta_{s}}}{l!s!2^l}{\cal
F}_{\alpha_1\alpha_2}\cdots{\cal
F}_{\alpha_{2l-1}\alpha_{2l}}\gamma_{\beta_1\ldots\beta_{s}} \ .
\eea
A key observation is that $\gamma_{\text{D}p}(\calf)^T\gamma_{\text{D}p}(\calf)=\bbone$ and thus $\gamma_{\text{D}p}(\calf)$ is an orthogonal matrix. This implies that static supersymmetric D-branes are allowed only if $\chi_1$ and $\chi_2$ have the same norm. We are then led to add the following necessary condition
for a D-calibrated background
\bea
\label{calnorm}
a^2=b^2\ .
\eea
{}From \eqref{kappa} follows furthermore that
\eq{
\label{calsource}
\mathbb{V} \cdot j_{(\Sigma,\calf)} = 0 \, .
}
Since $j_{(\Sigma,\calf)}$ can also be seen as the pure spinor defining the generalized tangent bundle associated to $(\Sigma,\calf)$
we find from the above equation that $v_{(+)}$ is along $\Sigma$ and
\eq{
v^{(-)}|_\Sigma = \iota_{v_{(+)}} \calf \, .
}
{}From the last equation in \eqref{9eq1} then follows that $\call_{v_{(+)}} \calf = 0$ so that $v_{(+)}$ is also a symmetry of the world-volume
gauge field on calibrated D-branes. Eqs.~\eqref{calnorm} and \eqref{calsource} can be combined as $\mathbb{V}^X \cdot j^X=0$
upon defining the ten-dimensional generalized vector $\mathbb{V}^X=(V_{(+)},V^{(-)})$ and using \eqref{extsource}.

It is easy to see that the D-brane supersymmetry condition \eqref{kappa} can be written in terms of a calibration condition of the form (\ref{calcond}) if we choose as calibration
\bea
\omega=\frac{16 e^{A-\Phi}}{a^2}\Psi\ .
\eea

It turns out that requiring that all sources are calibrated is not quite sufficient to ensure that $\omega$ is a proper calibration for probes.
Indeed, one has to check the two properties (\ref{alg2}) and (\ref{integr}). The algebraic condition (\ref{alg2}) is easily derived from the orthogonality of $\gamma_{\text{D}p}(\calf)$. The differential condition (\ref{integr}) on the other hand should follow from the background supersymmetry. However, from  (\ref{int1}) we see that (\ref{integr}) can be satisfied only if
\bea\label{cons}
v^{(-)}\wedge \hat F+\iota_{v_{(+)}}\hat F=0\ .
\eea
Thus it seems that in order to have a D-calibrated vacuum, we need to impose constraints involving $v_{(+)}$ and $v_{(-)}$.
The easiest way to impose then is to demand
\bea
\iota_{v_{(+)}}\hat F=0\quad\text{and}\quad v^{(-)}=0\ ,
\eea
which is indeed satisfied for the four-dimensional compactifications to Minkowski and AdS backgrounds.
It would however be interesting to see if there are other explicit  cases in which the condition (\ref{cons}) can be satisfied by some other means.\footnote{For example one might have non-identically vanishing $v^{(-)}$ but $\hat F=v^{(-)}\wedge \ldots$.}


\section{Generalized Hitchin flow}
\label{genhitchin}

Using the Poincar\'e coordinates for AdS$_4$ it is natural to define
\eq{
\rho = e^A \d z \wedge \Re \Psi_1 + \Re (e^{i\theta} \Psi_2) \, .
}
It follows that
\eq{
\hat{\rho}=\star_7 \sigma(\rho) = -e^A \d z \wedge \Im (e^{i\theta} \Psi_2) - \Im \Psi_1 \, ,
}
and thus
\begin{subequations}
\label{g2g2}
\begin{align}
\Theta^{\text{sf}}_P & = e^{3A+\frac{3z}{R}} \d t \wedge \d x^1 \wedge \d x^2 \wedge \rho \, , \\
\Theta^{\text{DW}}_P & = - e^{2A+\frac{2z}{R}} \, \d t \wedge \d x^1 \wedge \hat{\rho} \, .
\end{align}
\end{subequations}
In fact, $(\rho,\hat{\rho})$ are the pure spinors of a generalized $G_2 \times G_2$-structure. We can also find
them by defining seven-dimensional spinors
\begin{subequations}
\begin{align}
\eta^{(1)} & = e^{\frac{z}{2R}} \left[\left(\begin{array}{c} 1 \\ 0 \end{array} \right) \otimes \eta^{(1)}_+  \pm
e^{-i\theta} \left(\begin{array}{c} 0 \\ 1 \end{array} \right)\otimes \eta^{(1)}_-  \right]\, , \\
\eta^{(2)} & = e^{\frac{z}{2R}} \left[ \left(\begin{array}{c} 1 \\ 0 \end{array} \right) \otimes \eta^{(2)}_\mp \pm
e^{-i\theta} \left(\begin{array}{c} 0 \\ 1 \end{array} \right) \otimes\eta^{(2)}_\pm \right]\, ,
\end{align}
\end{subequations}
with the upper/lower sign for IIA/IIB respectively and the seven-dimensional gamma-matrices
\eq{
\gamma_i =  \sigma_3 \otimes \hat{\gamma}_i \, , \qquad \gamma_z = e^A   \sigma_1 \otimes \bbone \, .
}
Note that the ten-dimensional supersymmetry ansatz \eqref{adskilling} in terms of these seven-dimensional spinors becomes
\bea\label{g2killing}
\epsilon_1 & = & \zeta_0 \otimes \eta^{(1)}\, ,\cr
\epsilon_2 & = & \zeta_0 \otimes \eta^{(2)}\, ,
\eea
with $\zeta_0$ a constant three-dimensional spinor. Then we can write $(\rho,\hat{\rho})$ as bilinears of the
seven-dimensional spinors
\begin{subequations}
\begin{align}
\eta^{(1)} \eta^{(2)\dagger} & = \frac{||\eta^{(1)}|| ||\eta^{(2)}||}{8} \, \hat{\rho} \, , \\
(\bbone \otimes \sigma_2) \eta^{(1)} \eta^{(2)\dagger} & = - \frac{||\eta^{(1)}|| ||\eta^{(2)}||}{8} \rho \, .
\end{align}
\end{subequations}

{}From the general results of appendix \ref{susycalgen} (see also \cite{wittG2} in the absence of a warp factor and RR-fields)
we find that supersymmetry in compactifications of type II on $\mathbb{R}^{1,2}$ leads to a $G_2 \times G_2$-structure that satisfies
\eq{
\d_H(e^{3A +\frac{3z}{R}} \rho ) = - e^{4A +\frac{3z}{R}} \d z \wedge \tilde{F} \, , \qquad \d_H \left( e^{2A + \frac{2z}{R}} \hat{\rho} \right)= 0 \, .
}

For supersymmetric three-dimensional space-time filling D-branes
we find from the $\kappa$-condition $\hat{\Gamma}_{\text{D}p} \epsilon_2 = \epsilon_1$
\eq{
- (\sigma_2 \otimes \bbone) \hat{\gamma}_{(p-2)} \eta^{(2)} = \eta^{(1)} \, ,
}
which implies
\eq{
\sqrt{\text{det}(g_7|_\Sigma + \calf)} \, \d^{p-2} \sigma = \rho|_\Sigma \wedge e^\calf \, ,
}
so that $\rho$ is the calibration form. An alternative formulation of the
condition for the D-brane to be calibrated is
\eq{
\langle \mathbb{X} \cdot \hat{\rho} , j_{(\Sigma,\calf)} \rangle = 0 \, .
\label{g2cal}
}
Note that this F-flatness-type condition again implies the D-flatness.

\end{appendix}


\end{document}

%% file: boundaryfluccolor.tex
\begingroup\makeatletter%
\ifx\SetFigFont\undefined%
  \gdef\SetFigFont#1#2#3#4#5{%
    \reset@font\fontsize{#1}{#2pt}%
    \fontfamily{#3}\fontseries{#4}\fontshape{#5}%
    \selectfont}%
\fi
\endgroup%
\begin{pspicture}(-2.5,-0.0000)(13.6407,7.1691)
\newrgbcolor{sigmacolor}{0 0.5 0}
\newrgbcolor{deformcolor}{0 0 0.5}
\psellipse[linewidth=0.0159,linecolor=sigmacolor](11.6,4.6673)(1.1663,1.5621)
\begin{psclip}{\pspolygon[fillstyle=none,linestyle=none](11.6,7)(13.6250,7)(13.6250,1.5)(11.6,1.5)}
\psellipse[linewidth=0.0476,linecolor=sigmacolor,fillstyle=none](11.6,4.6673)(1.1663,1.58)
\end{psclip}
\psellipse[linewidth=0.0476,linecolor=sigmacolor](2.4532,4.6673)(1.1663,1.5621)
\psellipse[linewidth=0.0159,linecolor=deformcolor](5.0736,4.4027)(1.3,2.3580)
\psline[linewidth=0.0159,linecolor=sigmacolor](2.4532,6.2294)(11.6,6.2294)
\psline[linewidth=0.0159,linecolor=sigmacolor](2.4532,3.1052)(11.6,3.1052)
\psline[linewidth=0.0159,linecolor=black](2.3749,0.9377)(11.5549,0.9377)
\pspolygon[linewidth=0.0159,fillstyle=solid,fillcolor=black](2.3749,0.9377)(2.8829,0.8107)(2.8829,1.0647)
\pspolygon[linewidth=0.0159,fillstyle=solid,fillcolor=black](11.5549,0.9377)(11.0469,1.0647)(11.0469,0.8107)
\psline[linewidth=0.0159,linecolor=black](0.1524,6.6)(0.1524,2)
\pspolygon[linewidth=0.0159,fillstyle=solid,fillcolor=black](0.1524,7)(0.0254,6.5)(0.2794,6.5)
\pspolygon[linewidth=0.0159,fillstyle=solid,fillcolor=black](0.1524,2)(0.2794,2.5)(0.0254,2.5)
\pstVerb{2 setlinejoin}%
\psline[linewidth=0.0476,linecolor=deformcolor]{-}(2.4532,6.2294)(2.5146,6.2294)(2.5315,6.2369)(2.5591,6.2411)
	(2.6014,6.2553)(2.6522,6.260)(2.7114,6.2759)(2.7750,6.2844)
	(2.8384,6.2907)(2.8998,6.2971)(2.9570,6.3034)(3.0099,6.3098)
	(3.0565,6.3161)(3.1030,6.3204)(3.1433,6.3246)(3.1835,6.3310)
	(3.2195,6.3352)(3.2576,6.3394)(3.2935,6.3437)(3.3295,6.3500)
	(3.3655,6.3542)(3.4036,6.3606)(3.4438,6.3648)(3.4840,6.3712)
	(3.5242,6.3775)(3.5645,6.3860)(3.6068,6.3923)(3.6491,6.4008)
	(3.6915,6.4093)(3.7317,6.4177)(3.7740,6.4262)(3.8142,6.4347)
	(3.8545,6.4431)(3.8926,6.4516)(3.9307,6.4622)(3.9709,6.4707)
	(4.0090,6.4812)(4.0513,6.4939)(4.0915,6.5045)(4.1360,6.5172)
	(4.1804,6.5299)(4.2270,6.5426)(4.2736,6.5574)(4.3222,6.5722)
	(4.3709,6.5871)(4.4175,6.6019)(4.4662,6.6167)(4.5127,6.6315)
	(4.5572,6.6463)(4.6016,6.6590)(4.6440,6.6717)(4.6842,6.6844)
	(4.7223,6.6971)(4.7604,6.7098)(4.7964,6.7204)(4.8345,6.7310)
	(4.8704,6.7416)(4.9086,6.7522)(4.9445,6.7649)(4.9848,6.7755)
	(5.0229,6.7860)(5.0631,6.7945)(5.1033,6.8051)(5.1435,6.8157)
	(5.1837,6.8241)(5.2218,6.8326)(5.2620,6.8411)(5.3022,6.8495)
	(5.3403,6.8580)(5.3785,6.8644)(5.4187,6.8707)(5.4568,6.8771)
	(5.4928,6.8813)(5.5309,6.8876)(5.5690,6.8919)(5.6092,6.8982)
	(5.6515,6.9025)(5.6960,6.9088)(5.7404,6.9130)(5.7870,6.9194)
	(5.8335,6.9236)(5.8801,6.9300)(5.9288,6.9342)(5.9754,6.9406)
	(6.0219,6.9448)(6.0664,6.9490)(6.1108,6.9554)(6.1532,6.9596)
	(6.1934,6.9638)(6.2336,6.9681)(6.2717,6.9744)(6.3077,6.9787)
	(6.3479,6.9850)(6.3881,6.9892)(6.4262,6.9956)(6.4664,7.0019)
	(6.5066,7.0083)(6.5469,7.0146)(6.5892,7.0210)(6.6294,7.0273)
	(6.6696,7.0337)(6.7119,7.0400)(6.7522,7.0464)(6.7903,7.0527)
	(6.8305,7.0570)(6.8665,7.0633)(6.9046,7.0676)(6.9406,7.0718)
	(6.9765,7.0760)(7.0146,7.0803)(7.0506,7.0824)(7.0887,7.0866)
	(7.1268,7.0887)(7.1691,7.0908)(7.2115,7.0930)(7.2559,7.0930)
	(7.3025,7.0951)(7.3491,7.0951)(7.3978,7.0972)(7.4486,7.0972)
	(7.4972,7.0972)(7.5480,7.0972)(7.5967,7.0951)(7.6454,7.0951)
	(7.6941,7.0930)(7.7428,7.0930)(7.7893,7.0908)(7.8380,7.0887)
	(7.8804,7.0866)(7.9248,7.0845)(7.9714,7.0824)(8.0179,7.0803)
	(8.0666,7.0760)(8.1153,7.0739)(8.1661,7.0697)(8.2169,7.0676)
	(8.2698,7.0633)(8.3206,7.0591)(8.3735,7.0549)(8.4243,7.0527)
	(8.4751,7.0485)(8.5217,7.0443)(8.5704,7.0400)(8.6148,7.0358)
	(8.6572,7.0316)(8.6995,7.0273)(8.7376,7.0231)(8.7778,7.0189)
	(8.8180,7.0146)(8.8583,7.0083)(8.8985,7.0040)(8.9387,6.9977)
	(8.9789,6.9935)(9.0212,6.9871)(9.0615,6.9808)(9.1038,6.9744)
	(9.1440,6.9681)(9.1842,6.9617)(9.2265,6.9554)(9.2668,6.9490)
	(9.3049,6.9427)(9.3430,6.9363)(9.3832,6.9300)(9.4192,6.9257)
	(9.4573,6.9194)(9.4954,6.9130)(9.5356,6.9067)(9.5758,6.9003)
	(9.6160,6.8940)(9.6605,6.8876)(9.7049,6.8792)(9.7536,6.8707)
	(9.8002,6.8644)(9.8489,6.8538)(9.8996,6.8453)(9.9483,6.8368)
	(9.9949,6.8262)(10.0436,6.8178)(10.0880,6.8072)(10.1325,6.7966)
	(10.1727,6.7860)(10.2129,6.7755)(10.2489,6.7649)(10.2849,6.7522)
	(10.3251,6.7395)(10.3611,6.7268)(10.3992,6.7119)(10.4352,6.6950)
	(10.4712,6.6781)(10.5071,6.6612)(10.5431,6.6442)(10.5770,6.6252)
	(10.6109,6.6082)(10.6405,6.5913)(10.6722,6.5744)(10.6998,6.5574)
	(10.7273,6.5426)(10.7527,6.5278)(10.7781,6.5151)(10.8014,6.5024)
	(10.8331,6.4855)(10.8627,6.4707)(10.8966,6.4580)(10.9284,6.4431)
	(10.9622,6.4304)(10.9940,6.4198)(11.0278,6.4071)(11.0596,6.3987)
	(11.0892,6.3881)(11.1167,6.3796)(11.1421,6.3733)(11.1675,6.3648)
	(11.1951,6.3563)(11.2247,6.3479)(11.2564,6.3394)(11.2924,6.3288)
	(11.3305,6.3161)(11.3707,6.3056)(11.4088,6.2950)(11.4342,6.2865)
	(11.4469,6.2823)(11.4491,6.2823)(11.55,6.2294)(11.6,6.2294)
\pstVerb{0 setlinejoin}%
\pstVerb{2 setlinejoin}%
\psline[linewidth=0.0476,linecolor=deformcolor]{-}(2.4532,3.1052)(2.4553,3.1052)(2.4638,3.09)(2.4871,3.093)
	(2.5231,3.08)(2.5717,3.0607)(2.6268,3.0522)(2.6839,3.0438)
	(2.7390,3.0353)(2.7876,3.0268)(2.8342,3.0184)(2.8744,3.0099)
	(2.9125,3.0036)(2.9485,2.9951)(2.9824,2.9887)(3.0120,2.9803)
	(3.0438,2.9739)(3.0755,2.9655)(3.1073,2.9570)(3.1411,2.9485)
	(3.1750,2.9379)(3.2110,2.9274)(3.2470,2.9168)(3.2808,2.9041)
	(3.3168,2.8914)(3.3507,2.8787)(3.3845,2.8660)(3.4163,2.8533)
	(3.4480,2.8406)(3.4777,2.8258)(3.5073,2.8109)(3.5370,2.7961)
	(3.5666,2.7813)(3.5983,2.7644)(3.6301,2.7474)(3.6618,2.7284)
	(3.6936,2.7093)(3.7275,2.6882)(3.7613,2.6670)(3.7952,2.6458)
	(3.8291,2.6268)(3.8608,2.6056)(3.8904,2.5845)(3.9222,2.5654)
	(3.9497,2.5463)(3.9793,2.5294)(4.0069,2.5104)(4.0344,2.4934)
	(4.0619,2.4765)(4.0894,2.4596)(4.1190,2.4405)(4.1487,2.4236)
	(4.1783,2.4045)(4.2100,2.3876)(4.2418,2.3685)(4.2736,2.3516)
	(4.3053,2.3347)(4.3371,2.3177)(4.3688,2.3029)(4.4006,2.2881)
	(4.4302,2.2733)(4.4598,2.2585)(4.4916,2.2458)(4.5212,2.2331)
	(4.5530,2.2204)(4.5868,2.2077)(4.6207,2.1950)(4.6567,2.1823)
	(4.6948,2.1675)(4.7350,2.1527)(4.7752,2.1400)(4.8154,2.1251)
	(4.8556,2.1103)(4.8980,2.0955)(4.9382,2.0807)(4.9763,2.0659)
	(5.0144,2.0511)(5.0525,2.0362)(5.0906,2.0214)(5.1245,2.0087)
	(5.1604,1.9939)(5.1964,1.9791)(5.2324,1.9643)(5.2705,1.9473)
	(5.3107,1.9304)(5.3530,1.9135)(5.3975,1.8944)(5.4398,1.8775)
	(5.4864,1.8584)(5.5309,1.8415)(5.5753,1.8246)(5.6219,1.8076)
	(5.6663,1.7907)(5.7108,1.7759)(5.7552,1.7611)(5.7997,1.7484)
	(5.8441,1.7357)(5.8865,1.7230)(5.9288,1.7124)(5.9711,1.7018)
	(6.0177,1.6912)(6.0643,1.6806)(6.1129,1.6701)(6.1616,1.6595)
	(6.2145,1.6489)(6.2653,1.6404)(6.3182,1.6298)(6.3733,1.6214)
	(6.4262,1.6129)(6.4770,1.6044)(6.5278,1.5960)(6.5786,1.5896)
	(6.6273,1.5833)(6.6760,1.5769)(6.7204,1.5706)(6.7670,1.5642)
	(6.8093,1.5579)(6.8538,1.5536)(6.8961,1.5473)(6.9406,1.5431)
	(6.9850,1.5388)(7.0316,1.5346)(7.0760,1.5282)(7.1247,1.5261)
	(7.1713,1.5219)(7.2178,1.5177)(7.2665,1.5155)(7.3152,1.5134)
	(7.3618,1.5113)(7.4083,1.5113)(7.4549,1.5113)(7.4994,1.5113)
	(7.5438,1.5134)(7.5882,1.5134)(7.6306,1.5177)(7.6729,1.5198)
	(7.7153,1.5240)(7.7555,1.5282)(7.7999,1.5346)(7.8422,1.5409)
	(7.8867,1.5473)(7.9333,1.5558)(7.9798,1.5642)(8.0285,1.5748)
	(8.0751,1.5854)(8.1238,1.5981)(8.1724,1.6108)(8.2190,1.6235)
	(8.2656,1.6362)(8.3122,1.6510)(8.3566,1.6658)(8.3968,1.6785)
	(8.4370,1.6933)(8.4751,1.7082)(8.5111,1.7230)(8.5471,1.7378)
	(8.5788,1.7526)(8.6191,1.7716)(8.6572,1.7928)(8.6931,1.8140)
	(8.7312,1.8352)(8.7672,1.8584)(8.8053,1.8838)(8.8413,1.9092)
	(8.8773,1.9346)(8.9112,1.9600)(8.9450,1.9875)(8.9789,2.0130)
	(9.0107,2.0405)(9.0424,2.0659)(9.0741,2.0913)(9.1038,2.1167)
	(9.1355,2.1400)(9.1631,2.1632)(9.1927,2.1865)(9.2223,2.2098)
	(9.2541,2.2331)(9.2858,2.2585)(9.3218,2.2839)(9.3578,2.3093)
	(9.3938,2.3347)(9.4319,2.3622)(9.4679,2.3876)(9.5060,2.4130)
	(9.5441,2.4384)(9.5822,2.4617)(9.6181,2.4850)(9.6541,2.5061)
	(9.6880,2.5252)(9.7218,2.5463)(9.7557,2.5633)(9.7896,2.5823)
	(9.8235,2.5993)(9.8594,2.6162)(9.8954,2.6331)(9.9314,2.6501)
	(9.9695,2.6670)(10.0076,2.6839)(10.0457,2.7009)(10.0838,2.7157)
	(10.1219,2.7305)(10.1600,2.7453)(10.1981,2.7601)(10.2341,2.7728)
	(10.2680,2.7855)(10.3018,2.7961)(10.3336,2.8088)(10.3653,2.8194)
	(10.3971,2.8300)(10.4309,2.8406)(10.4648,2.8512)(10.4987,2.8638)
	(10.5346,2.8744)(10.5685,2.8850)(10.6066,2.8956)(10.6426,2.9083)
	(10.6786,2.9189)(10.7146,2.9295)(10.7506,2.9379)(10.7865,2.9485)
	(10.8183,2.9570)(10.8522,2.9655)(10.8839,2.9739)(10.9135,2.9803)
	(10.9432,2.9887)(10.9770,2.9951)(11.0109,3.0036)(11.0469,3.0099)
	(11.0871,3.0184)(11.1294,3.0268)(11.1781,3.0353)(11.2289,3.0438)
	(11.2839,3.0522)(11.3369,3.0607)(11.3813,3.0692)(11.4173,3.0755)
	(11.4385,3.0776)(11.4469,3.0797)(11.4491,3.0797)(11.55,3.1052)(11.6,3.1052)
\pstVerb{0 setlinejoin}%
\pstVerb{2 setlinejoin}%
\psline[linewidth=0.0159,linecolor=deformcolor]{-}(6.3966,2.0489)(6.3987,2.0489)(6.4050,2.0468)(6.4220,2.0426)
	(6.4516,2.0362)(6.4918,2.0278)(6.5405,2.0172)(6.5934,2.0045)
	(6.6463,1.9939)(6.6993,1.9833)(6.7458,1.9727)(6.7903,1.9643)
	(6.8305,1.9558)(6.8665,1.9494)(6.8982,1.9431)(6.9300,1.9389)
	(6.9617,1.9325)(6.9956,1.9283)(7.0295,1.9240)(7.0654,1.9198)
	(7.1014,1.9156)(7.1395,1.9135)(7.1776,1.9113)(7.2136,1.9092)
	(7.2517,1.9092)(7.2877,1.9092)(7.3237,1.9092)(7.3575,1.9113)
	(7.3914,1.9135)(7.4231,1.9177)(7.4549,1.9198)(7.4866,1.9262)
	(7.5184,1.9325)(7.5523,1.9389)(7.5882,1.9473)(7.6306,1.9579)
	(7.6750,1.9706)(7.7237,1.9854)(7.7745,2.0002)(7.8232,2.0151)
	(7.8677,2.0299)(7.9015,2.0383)(7.9206,2.0468)(7.9290,2.0489)
	(7.9312,2.0489)
\pstVerb{0 setlinejoin}%
\pstVerb{2 setlinejoin}%
\psline[linewidth=0.0159,linecolor=deformcolor]{-}(6.6082,2.2860)(6.6104,2.2860)(6.6167,2.2839)(6.6379,2.2818)
	(6.6675,2.2754)(6.7098,2.2691)(6.7564,2.2606)(6.8051,2.2521)
	(6.8516,2.2458)(6.8940,2.2373)(6.9342,2.2331)(6.9702,2.2267)
	(7.0019,2.2225)(7.0337,2.2183)(7.0633,2.2140)(7.0930,2.2119)
	(7.1247,2.2077)(7.1564,2.2056)(7.1903,2.2035)(7.2242,2.2013)
	(7.2602,2.1992)(7.2961,2.1971)(7.3300,2.1971)(7.3639,2.1950)
	(7.3978,2.1950)(7.4295,2.1971)(7.4612,2.1971)(7.4888,2.1992)
	(7.5184,2.2013)(7.5502,2.2056)(7.5819,2.2098)(7.6158,2.2162)
	(7.6539,2.2246)(7.6941,2.2331)(7.7406,2.2437)(7.7872,2.2564)
	(7.8317,2.2670)(7.8677,2.2775)(7.8909,2.2818)(7.9015,2.2860)
	(7.9036,2.2860)
\pstVerb{0 setlinejoin}%
\pstVerb{2 setlinejoin}%
\psline[linewidth=0.0159,linecolor=deformcolor]{-}(6.6865,6.7839)(6.6887,6.7839)(6.6993,6.7860)(6.7225,6.7924)
	(6.7628,6.8009)(6.8136,6.8114)(6.8728,6.8262)(6.9321,6.8389)
	(6.9871,6.8495)(7.0379,6.8601)(7.0824,6.8707)(7.1226,6.8792)
	(7.1564,6.8855)(7.1861,6.8898)(7.2157,6.8940)(7.2475,6.9003)
	(7.2771,6.9046)(7.3088,6.9067)(7.3385,6.9088)(7.3702,6.9109)
	(7.4020,6.9130)(7.4316,6.9130)(7.4634,6.9130)(7.4930,6.9109)
	(7.5248,6.9088)(7.5544,6.9067)(7.5861,6.9046)(7.6158,6.9003)
	(7.6454,6.8982)(7.6793,6.8940)(7.7195,6.8898)(7.7639,6.8834)
	(7.8147,6.8771)(7.8698,6.8686)(7.9290,6.8622)(7.9883,6.8538)
	(8.0391,6.8474)(8.0793,6.8411)(8.1026,6.8389)(8.1132,6.8368)
	(8.1153,6.8368)
\pstVerb{0 setlinejoin}%
\pstVerb{2 setlinejoin}%
\psline[linewidth=0.0159,linecolor=deformcolor]{-}(6.8559,6.6188)(6.8580,6.6188)(6.8686,6.6231)(6.8940,6.6294)
	(6.9321,6.6400)(6.9765,6.6527)(7.0252,6.6675)(7.0718,6.6802)
	(7.1120,6.6908)(7.1480,6.6993)(7.1776,6.7077)(7.2051,6.7119)
	(7.2327,6.7183)(7.2559,6.7225)(7.2813,6.7246)(7.3067,6.7289)
	(7.3343,6.7310)(7.3597,6.7331)(7.3872,6.7331)(7.4147,6.7331)
	(7.4443,6.7331)(7.4718,6.7331)(7.4994,6.7310)(7.5248,6.7289)
	(7.5544,6.7268)(7.5777,6.7246)(7.6052,6.7204)(7.6327,6.7183)
	(7.6666,6.7141)(7.7047,6.7098)(7.7449,6.7035)(7.7915,6.6993)
	(7.8401,6.6929)(7.8888,6.6865)(7.9312,6.6802)(7.9629,6.6760)
	(7.9820,6.6738)(7.9904,6.6717)(7.9925,6.6717)
\pstVerb{0 setlinejoin}%
\begingroup\SetFigFont{15}{24.0}{\rmdefault}{\mddefault}{\updefault}%
\rput[lb](-2,6.95){(a)}%
\endgroup%
\begingroup\SetFigFont{15}{24.0}{\rmdefault}{\mddefault}{\updefault}%
\rput[lb](6.3437,0.1){AdS$_4$}%
\endgroup%
\begingroup\SetFigFont{15}{24.0}{\rmdefault}{\mddefault}{\updefault}%
\rput[lb](0.0318,1.4){internal}%
\endgroup%
\begingroup\SetFigFont{15}{24.0}{\rmdefault}{\mddefault}{\updefault}%
\rput[lb](1.1176,6.0018){\sigmacolor$\Sigma$}%
\endgroup%
\begingroup\SetFigFont{15}{24.0}{\rmdefault}{\mddefault}{\updefault}%
\rput[lb](4.05,6.95){\deformcolor$\Sigma'(\rho)$}%
\endgroup%
\end{pspicture}

%% file: homfluccolor.tex
\begingroup\makeatletter%
\ifx\SetFigFont\undefined%
  \gdef\SetFigFont#1#2#3#4#5{%
    \reset@font\fontsize{#1}{#2pt}%
    \fontfamily{#3}\fontseries{#4}\fontshape{#5}%
    \selectfont}%
\fi
\endgroup%
\begin{pspicture}(-2.5,-0.0000)(13.6250,6.8834)
\newrgbcolor{sigmacolor}{0 0.5 0}
\newrgbcolor{deformcolor}{0 0 0.5}
\newrgbcolor{shadecolor}{0.5 0 0}
\newrgbcolor{shadecolorlight}{1 0.3 0.3}
\psellipse[linewidth=0.0476,linecolor=deformcolor,fillcolor=shadecolorlight,hatchcolor=shadecolor,fillstyle=hlines*,hatchangle=45.00](2.7707,4.3117)(1.8013,2.5146)
\psellipse[linewidth=0.0159,linecolor=deformcolor,fillcolor=shadecolorlight,hatchcolor=shadecolor,fillstyle=hlines*,hatchangle=45.00](11.7729,4.3117)(1.8013,2.5146)
\psellipse[linewidth=0.0159,linecolor=sigmacolor,fillstyle=solid,fillcolor=white](11.7729,4.3117)(1.0604,1.5621)
\psellipse[linewidth=0.0476,linecolor=sigmacolor,fillstyle=solid,fillcolor=white](2.7707,4.3117)(1.0604,1.5621)
\psline[linewidth=0.0476,linecolor=sigmacolor]{-}(2.7982,5.8738)(4.1,5.8738)
\psline[linewidth=0.0159,linecolor=sigmacolor]{-}(4.1,5.8738)(11.7729,5.8738)
\psline[linewidth=0.0159,linecolor=sigmacolor]{-}(4.1,2.7496)(11.7729,2.7496)
\psline[linewidth=0.0476,linecolor=sigmacolor]{-}(2.7982,2.7496)(4.1,2.7496)
\psline[linewidth=0.0159,linecolor=black,fillcolor=white,hatchcolor=black,fillstyle=hlines*,hatchangle=45.00]{-}(0.1524,6.6527)(0.1524,1.8627)
\begin{psclip}{\pspolygon[fillstyle=none,linestyle=none](11.7729,7)(13.6250,7)(13.6250,1.5)(11.7729,1.5)}
\psellipse[linewidth=0.0476,linecolor=deformcolor,fillstyle=none](11.7729,4.3117)(1.8013,2.5146)
\end{psclip}
\psline[linewidth=0.0476,linecolor=deformcolor]{-}(2.7707,6.8263)(11.7729,6.8263)
\psline[linewidth=0.0476,linecolor=deformcolor]{-}(2.7707,1.7971)(11.7729,1.7971)
\psline[linewidth=0.0159,linecolor=black]{-}(2.5866,0.8848)(11.7729,0.8848)
\pspolygon[linewidth=0.0159,fillstyle=solid,fillcolor=black](2.5866,0.8848)(3.0946,0.7578)(3.0946,1.0118)
\pspolygon[linewidth=0.0159,fillstyle=solid,fillcolor=black](11.7666,0.8848)(11.2586,1.0118)(11.2586,0.7578)
\pspolygon[linewidth=0.0159,fillstyle=solid,fillcolor=black](0.1524,6.6527)(0.0254,6.1447)(0.2794,6.1447)
\pspolygon[linewidth=0.0159,fillstyle=solid,fillcolor=black](0.1524,1.8627)(0.2794,2.3707)(0.0254,2.3707)
\begingroup\SetFigFont{15}{24.0}{\rmdefault}{\mddefault}{\updefault}%
\rput[lb](-2,6.5){(b)}%
\endgroup%
\begingroup\SetFigFont{15}{24.0}{\rmdefault}{\mddefault}{\updefault}%
\rput[lb](6.3690,0.1222){AdS$_4$}%
\endgroup%
\begingroup\SetFigFont{15}{24.0}{\rmdefault}{\mddefault}{\updefault}%
\rput[lb](3.0,4.4831){$\sigmacolor\Sigma$}%
\endgroup%
\begingroup\SetFigFont{15}{24.0}{\rmdefault}{\mddefault}{\updefault}%
\rput[lb](4.6,5.0906){$\deformcolor\Sigma'$}%
\endgroup%
\begingroup\SetFigFont{15}{24.0}{\rmdefault}{\mddefault}{\updefault}%
\rput[lb](0.0318,1.2){internal}%
\endgroup%
\begingroup\SetFigFont{15}{24.0}{\rmdefault}{\mddefault}{\updefault}%
\rput[lb](7,6.1){${\cal B}$}%
\endgroup%
\end{pspicture}

%% file: adsbranefinal5.bbl
\begin{thebibliography}{99}

\bibitem{kklt}
  S.~Kachru, R.~Kallosh, A.~Linde and S.~P.~Trivedi,
  ``De Sitter vacua in string theory,''
  Phys.\ Rev.\  D {\bf 68} (2003) 046005
  [arXiv:hep-th/0301240].

\bibitem{effective}
  P.~Koerber and L.~Martucci,
  ``From ten to four and back again: how to generalize the geometry,''
  JHEP {\bf 0708} (2007) 059 [arXiv:0707.1038 [hep-th]].

\bibitem{zwirner}
  J.~P.~Derendinger, C.~Kounnas, P.~M.~Petropoulos and F.~Zwirner,
  ``Superpotentials in IIA compactifications with general fluxes,''
  Nucl.\ Phys.\  B {\bf 715} (2005) 211
  [arXiv:hep-th/0411276].
  G.~Villadoro and F.~Zwirner,
  ``$\caln = 1$ effective potential from dual type-IIA D6/O6 orientifolds with
  general fluxes,''
  JHEP {\bf 0506} (2005) 047
  [arXiv:hep-th/0503169].


\bibitem{dewolfe}
  O.~DeWolfe, A.~Giryavets, S.~Kachru and W.~Taylor,
  ``Type IIA moduli stabilization,''
  JHEP {\bf 0507} (2005) 066
  [arXiv:hep-th/0505160].


\bibitem{palti}
  T.~House and E.~Palti,
  ``Effective action of (massive) IIA on manifolds with $SU(3)$ structure,''
  Phys.\ Rev.\  D {\bf 72} (2005) 026004
  [arXiv:hep-th/0505177].

\bibitem{camara}
  P.~G.~C\'amara, A.~Font and L.~E.~Ib\'a\~{n}ez,
  ``Fluxes, moduli fixing and MSSM-like vacua in a simple IIA orientifold,''
  JHEP {\bf 0509} (2005) 013
  [arXiv:hep-th/0506066].


\bibitem{acharya}
  B.~S.~Acharya, F.~Benini and R.~Valandro,
  ``Fixing moduli in exact type IIA flux vacua,''
  JHEP {\bf 0702} (2007) 018
  [arXiv:hep-th/0607223].

\bibitem{gmpt}
  M.~Gra\~{n}a, R.~Minasian, M.~Petrini and A.~Tomasiello,
  ``Generalized structures of $\caln=1$ vacua,''
  JHEP {\bf 0511} (2005) 020
  [arXiv:hep-th/0505212].

\bibitem{hitchin}
  N.~Hitchin,
  ``Generalized Calabi-Yau manifolds,''
  Quart.\ J.\ Math.\ Oxford Ser.\  {\bf 54} (2003) 281
  [arXiv:math.dg/0209099].

\bibitem{gualtieri}
  M.~Gualtieri,
  ``Generalized complex geometry,''
  arXiv:math.dg/0401221.

\bibitem{tomasiello}
  A.~Tomasiello,
  ``Reformulating supersymmetry with a generalized Dolbeault operator,''
  arXiv:0704.2613 [hep-th].

\bibitem{harvey}
  R.~Harvey and H.~B.~Lawson,
  ``Calibrated geometries,''
  Acta Math.\  {\bf 148} (1982) 47.

\bibitem{papa}
  J.~Gutowski and G.~Papadopoulos,
  ``AdS calibrations,''
  Phys.\ Lett.\  B {\bf 462} (1999) 81
  [arXiv:hep-th/9902034];
  J.~Gutowski, G.~Papadopoulos and P.~K.~Townsend,
  ``Supersymmetry and generalized calibrations,''
  Phys.\ Rev.\  D {\bf 60} (1999) 106006
  [arXiv:hep-th/9905156].

\bibitem{gencal}
  P.~Koerber,
  ``Stable D-branes, calibrations and generalized Calabi-Yau geometry,''
  JHEP {\bf 0508} (2005) 099
  [arXiv:hep-th/0506154].

\bibitem{luca1}
  L.~Martucci and P.~Smyth,
  ``Supersymmetric D-branes and calibrations on general $\caln=1$ backgrounds,''
  JHEP {\bf 0511} (2005) 048
  [arXiv:hep-th/0507099].

\bibitem{bf}
  P.~Breitenlohner and D.~Z.~Freedman,
  ``Positive energy in Anti-de Sitter backgrounds and gauged extended supergravity,''
  Phys.\ Lett.\  B {\bf 115} (1982) 197;
  P.~Breitenlohner and D.~Z.~Freedman,
  ``Stability in gauged extended supergravity,''
  Annals Phys.\  {\bf 144} (1982) 249.

\bibitem{freedman}
  C.~J.~C.~Burges, D.~Z.~Freedman, S.~Davis and G.~W.~Gibbons,
  ``Supersymmetry in Anti-de Sitter Space,''
  Annals Phys.\  {\bf 167} (1986) 285;
  D.~Z.~Freedman,
  ``A fond farewell to Anti-de Sitter space,''
  Nuffield Workshop 1985:341.

\bibitem{myers}
  R.~C.~Myers,
  ``Dielectric-branes,''
  JHEP {\bf 9912} (1999) 022
  [arXiv:hep-th/9910053].

\bibitem{acharyadenef}
  B.~S.~Acharya, F.~Denef, C.~Hofman and N.~Lambert,
  ``Freund-Rubin revisited,''
  arXiv:hep-th/0308046.

\bibitem{jarahluc}
  J.~Evslin and L.~Martucci,
  ``D-brane networks in flux vacua, generalized cycles and calibrations,''
  JHEP {\bf 0707} (2007) 040
  [arXiv:hep-th/0703129].

\bibitem{luca2}
  L.~Martucci,
  ``D-branes on general $\caln = 1$ backgrounds: superpotentials and D-terms,''
  JHEP {\bf 0606} (2006) 033
  [arXiv:hep-th/0602129].


\bibitem{kr}
  A.~Karch and L.~Randall,
  ``Open and closed string interpretation of SUSY CFT's on branes with boundaries,''
  JHEP {\bf 0106} (2001) 063
  [arXiv:hep-th/0105132].

\bibitem{mtaylor}
  K.~Skenderis and M.~Taylor,
  ``Branes in AdS and pp-wave spacetimes,''
  JHEP {\bf 0206} (2002) 025
  [arXiv:hep-th/0204054].

\bibitem{urangacasc}
  J.~F.~G.~Cascales and A.~M.~Uranga,
  ``Branes on generalized calibrated submanifolds,''
  JHEP {\bf 0411} (2004) 083
  [arXiv:hep-th/0407132].

\bibitem{democratic}
  E.~Bergshoeff, R.~Kallosh, T.~Ort\'{\i}n, D.~Roest and A.~Van Proeyen,
  ``New formulations of $D = 10$ supersymmetry and D8--O8 domain walls,''
  Class.\ Quant.\ Grav.\  {\bf 18} (2001) 3359
  [arXiv:hep-th/0103233].

\bibitem{granascan}
  M.~Gra\~{n}a, R.~Minasian, M.~Petrini and A.~Tomasiello,
  ``A scan for new $\caln=1$ vacua on twisted tori,''
  JHEP {\bf 0705} (2007) 031
  [arXiv:hep-th/0609124].

\bibitem{deforms}
  P.~Koerber and L.~Martucci,
  ``Deformations of calibrated D-branes in flux generalized complex manifolds,''
  JHEP {\bf 0612} (2006) 062
  [arXiv:hep-th/0610044].

\bibitem{lucanapoli}
  L.~Martucci,
  ``Supersymmetric D-branes on flux backgrounds,''
  Fortsch.\ Phys.\  {\bf 55} (2007) 771
  [arXiv:hep-th/0701093].

\bibitem{cvetic}
  K.~Behrndt and M.~Cveti\v{c},
  ``General $\caln=1$ supersymmetric flux vacua of (massive) type IIA string   theory,''
  Phys.\ Rev.\ Lett.\  {\bf 95} (2005) 021601
  [arXiv:hep-th/0403049];
  K.~Behrndt and M.~Cveti\v{c},
  ``General $\caln=1$ supersymmetric fluxes in massive type IIA string theory,''
  Nucl.\ Phys.\  B {\bf 708} (2005) 45
  [arXiv:hep-th/0407263].


\bibitem{tsimpis}
  D.~L\"ust and D.~Tsimpis,
  ``Supersymmetric AdS$_4$ compactifications of IIA supergravity,''
  JHEP {\bf 0502} (2005) 027
  [arXiv:hep-th/0412250].

\bibitem{dealwis}
  S.~P.~de Alwis,
  ``Potentials for light moduli in supergravity and string theory,''
  arXiv:hep-th/0602182.

\bibitem{dstring}
  G.~Dvali, R.~Kallosh and A.~Van Proeyen,
  ``D-term strings,''
  JHEP {\bf 0401} (2004) 035
  [arXiv:hep-th/0312005];
  P.~Bin\'etruy, G.~Dvali, R.~Kallosh and A.~Van Proeyen,
  ``Fayet-Iliopoulos terms in supergravity and cosmology,''
  Class.\ Quant.\ Grav.\  {\bf 21} (2004) 3137
  [arXiv:hep-th/0402046].

\bibitem{duff}
  M.~J.~Duff,
  ``Anti-de Sitter space, branes, singletons, superconformal field theories
  and all that,''
  Int.\ J.\ Mod.\ Phys.\  A {\bf 14} (1999) 815
  [arXiv:hep-th/9808100].

\bibitem{town}
  H.~L\"u, C.~N.~Pope and P.~K.~Townsend,
  ``Domain walls from anti-de Sitter spacetime,''
  Phys.\ Lett.\  B {\bf 391} (1997) 39
  [arXiv:hep-th/9607164].


\bibitem{tsimpisII}
  C.~Kounnas, D.~L\"ust, P.~M.~Petropoulos and D.~Tsimpis,
  ``AdS$_4$ flux vacua in type II superstrings and their domain-wall solutions,''
  JHEP {\bf 0709} (2007) 051
  [arXiv:0707.4270 [hep-th]].

\bibitem{chiossal}
  S.~Chiossi and S.~Salamon, ``The intrinsic torsion of $SU(3)$ and $G_2$-structures,''
  in Differential geometry, Valencia (2001), World Sci.\ Publishing, p.\ 115, [arXiv:math.DG/0202282].

\bibitem{paultsimpis}
  P.~Koerber and D.~Tsimpis,
  ``Supersymmetric sources, integrability and generalized-structure
  compactifications,''
  JHEP {\bf 0708} (2007) 082 [arXiv:0706.1244 [hep-th]].

\bibitem{nogo}
  B.~de Wit, D.~J.~Smit and N.~D.~Hari Dass,
  ``Residual supersymmetry of compactified $D=10$ supergravity,''
  Nucl.\ Phys.\  B {\bf 283} (1987) 165;
  J.~M.~Maldacena and C.~N\'u\~{n}ez,
  ``Supergravity description of field theories on curved manifolds and a no go theorem,''
  Int.\ J.\ Mod.\ Phys.\  A {\bf 16} (2001) 822
  [arXiv:hep-th/0007018].

\bibitem{beyondSE}
  J.~P.~Gauntlett, D.~Martelli, J.~Sparks and D.~Waldram,
  ``Supersymmetric AdS$_5$ solutions of type IIB supergravity,''
  Class.\ Quant.\ Grav.\  {\bf 23} (2006) 4693
  [arXiv:hep-th/0510125].

\bibitem{nearlyK}
  J.-B. Butruille,
  ``Homogeneous nearly K\"ahler manifolds,''
  arXiv:math/0612655.

\bibitem{baez}
  For an excellent review of the octonions see
  J. Baez, ``The octonions,''
  Bull.\ Amer.\ Math.\ Soc.\ {\bf 39} (2002), 145 (errata in Bull.\ Amer.\ Math.\ Soc.\ {\bf 42} (2005), 213)
  [arXiv:math/0105155].

\bibitem{ivanpap}
  J.~Gutowski, S.~Ivanov and G.~Papadopoulos,
  ``Deformations of generalized calibrations and compact non-K\"ahler manifolds
  with vanishing first Chern class,''
  Asian J.\ Math.\ {\bf 7} (2003) 39 [arXiv:math/0205012].

\bibitem{coisotropic}
  A.~Kapustin and D.~Orlov,
  ``Remarks on A-branes, mirror symmetry, and the Fukaya category,''
  J.\ Geom.\ Phys.\  {\bf 48} (2003) 84
  [arXiv:hep-th/0109098].

\bibitem{marchesanoco}
  A.~Font, L.~E.~Ibanez and F.~Marchesano,
  ``Coisotropic D8-branes and model-building,''
  JHEP {\bf 0609} (2006) 080
  [arXiv:hep-th/0607219].



\bibitem{granaN2}
  M.~Gra\~{n}a, J.~Louis and D.~Waldram,
  ``Hitchin functionals in $N = 2$ supergravity,''
  JHEP {\bf 0601} (2006) 008
  [arXiv:hep-th/0505264];
  M.~Gra\~{n}a, J.~Louis and D.~Waldram,
  ``$SU(3) \times SU(3)$ compactification and mirror duals of magnetic fluxes,''
  JHEP {\bf 0704} (2007) 101
  [arXiv:hep-th/0612237].

\bibitem{grimm}
  I.~Benmachiche and T.~W.~Grimm,
  ``Generalized $N=1$ orientifold compactifications and the Hitchin
  functionals,''
  Nucl.\ Phys.\  B {\bf 748} (2006) 200
  [arXiv:hep-th/0602241].

\bibitem{cassbil}
  D.~Cassani and A.~Bilal,
  ``Effective actions and $N=1$ vacuum conditions from $SU(3) \times SU(3)$
  compactifications,''
  JHEP {\bf 0709} (2007) 076
  [arXiv:0707.3125 [hep-th]].

\bibitem{kp}
  A.-K.~Kashani-Poor,
  ``Nearly K\"ahler Reduction,''
  JHEP {\bf 0711} (2007) 026 [arXiv:0709.4482 [hep-th]].

\bibitem{giantgraviton}
  J.~McGreevy, L.~Susskind and N.~Toumbas,
  ``Invasion of the giant gravitons from anti-de Sitter space,''
  JHEP {\bf 0006} (2000) 008
  [arXiv:hep-th/0003075];
  M.~T.~Grisaru, R.~C.~Myers and \O.~Tafjord,
  ``SUSY and Goliath,''
  JHEP {\bf 0008} (2000) 040
  [arXiv:hep-th/0008015].

\bibitem{bion}
  C.~G.~Callan and J.~M.~Maldacena,
  ``Brane dynamics from the Born-Infeld action,''
  Nucl.\ Phys.\  B {\bf 513} (1998) 198
  [arXiv:hep-th/9708147];
  G.~W.~Gibbons,
  ``Born-Infeld particles and Dirichlet p-branes,''
  Nucl.\ Phys.\  B {\bf 514} (1998) 603
  [arXiv:hep-th/9709027].

\bibitem{hjsmith}
  E.~J.~Hackett-Jones and D.~J.~Smith,
  ``Type IIB Killing spinors and calibrations,''
  JHEP {\bf 0411} (2004) 029
  [arXiv:hep-th/0405098].

\bibitem{KK}
  A.~Karch and E.~Katz,
  ``Adding flavor to AdS/CFT,''
  JHEP {\bf 0206} (2002) 043
  [arXiv:hep-th/0205236].

\bibitem{malda}
  O.~Aharony, A.~Fayyazuddin and J.~M.~Maldacena,
  ``The large $N$ limit of $N = 2,1$ field theories from three-branes in F-theory,''
  JHEP {\bf 9807} (1998) 013
  [arXiv:hep-th/9806159].

\bibitem{hashimoto}
  S.~A.~Cherkis and A.~Hashimoto,
  ``Supergravity solution of intersecting branes and AdS/CFT with flavor,''
  JHEP {\bf 0211} (2002) 036
  [arXiv:hep-th/0210105].

\bibitem{benini}
  F.~Benini, F.~Canoura, S.~Cremonesi, C.~N\'u\~{n}ez and A.~V.~Ramallo,
  ``Unquenched flavors in the Klebanov-Witten model,''
  JHEP {\bf 0702} (2007) 090
  [arXiv:hep-th/0612118].


\bibitem{calold}
  K.~Becker, M.~Becker and A.~Strominger,
  ``Five-branes, membranes and non-perturbative string theory,''
  Nucl.\ Phys.\  B {\bf 456} (1995) 130
  [arXiv:hep-th/9507158];
  K.~Becker, M.~Becker, D.~R.~Morrison, H.~Ooguri, Y.~Oz and Z.~Yin,
  ``Supersymmetric cycles in exceptional holonomy manifolds and Calabi-Yau 4-folds,''
  Nucl.\ Phys.\  B {\bf 480} (1996) 225
  [arXiv:hep-th/9608116];
  G.~W.~Gibbons and G.~Papadopoulos,
  ``Calibrations and intersecting branes,''
  Commun.\ Math.\ Phys.\  {\bf 202} (1999) 593
  [arXiv:hep-th/9803163].

\bibitem{gauntlettcalbulk}
  J.~P.~Gauntlett, N.-w.~Kim, D.~Martelli and D.~Waldram,
  ``Fivebranes wrapped on SLAG three-cycles and related geometry,''
  JHEP {\bf 0111} (2001) 018
  [arXiv:hep-th/0110034];
  J.~P.~Gauntlett, D.~Martelli, S.~Pakis and D.~Waldram,
  ``G-structures and wrapped NS5-branes,''
  Commun.\ Math.\ Phys.\  {\bf 247} (2004) 421
  [arXiv:hep-th/0205050];
  J.~P.~Gauntlett, D.~Martelli and D.~Waldram,
  ``Superstrings with intrinsic torsion,''
  Phys.\ Rev.\  D {\bf 69} (2004) 086002
  [arXiv:hep-th/0302158];
  D.~Martelli and J.~Sparks,
  ``G-structures, fluxes and calibrations in M-theory,''
  Phys.\ Rev.\  D {\bf 68} (2003) 085014
  [arXiv:hep-th/0306225].

\bibitem{lucadirac}
  L.~Martucci, J.~Rosseel, D.~Van den Bleeken and A.~Van Proeyen,
  ``Dirac actions for D-branes on backgrounds with fluxes,''
  Class.\ Quant.\ Grav.\  {\bf 22} (2005) 2745
  [arXiv:hep-th/0504041].

\bibitem{superalgebra}
  J.~Figueroa-O'Farrill, E.~Hackett-Jones and G.~Moutsopoulos,
  ``The Killing superalgebra of ten-dimensional supergravity backgrounds,''
  Class.\ Quant.\ Grav.\  {\bf 24} (2007) 3291
  [arXiv:hep-th/0703192].

\bibitem{wittG2}
  F.~Witt,
  ``Generalised $G_2$-manifolds,''
  Commun.\ Math.\ Phys.\  {\bf 265} (2006) 275
  [arXiv:math.dg/0411642];
  C.~Jeschek and F.~Witt,
  ``Generalised $G_2$-structures and type IIB superstrings,''
  JHEP {\bf 0503} (2005) 053
  [arXiv:hep-th/0412280].

\end{thebibliography}
